\documentclass[useAMS,usenatbib]{mn2e}
\usepackage{epsfig}
\usepackage{amssymb}

\def\araa{{\it Annu. Rev. Astron. Astrophys.} \,}

\def\apj{{\it ApJ \,}}

\def\apjl{{\it Ap. J. Lett.} \,}

\def\mnras{{\it MNRAS} \,}

\def\nat{{\it Nature} \,}
\def\aap{{\it A\&A} \,}
\def\araa{{\it } \,}

\title[A1689 Mass and Gas Profiles]{\it Mass and Gas Profiles in A1689: Joint X-ray and Lensing 
Analysis}
\author[Doron Lemze et al.]{Doron Lemze$^{1}$, Rennan Barkana$^{1}$, Tom J.\ Broadhurst$^{1}$ 
\& Yoel Rephaeli$^{1}$\\
$^{1}$School of Physics and Astronomy, Tel Aviv University, Tel Aviv, 69978, 
Israel\\}

\begin{document}

\pagerange{\pageref{firstpage}--\pageref{lastpage}} \pubyear{2007}

\maketitle

\label{firstpage}

\begin{abstract}

We carry out a comprehensive joint analysis of high quality HST/ACS
and Chandra measurements of A1689, from which we derive mass,
temperature, X-ray emission and abundance profiles. The X-ray emission
is smooth and symmetric, and the lensing mass is centrally
concentrated indicating a relaxed cluster. Assuming hydrostatic
equilibrium we deduce a 3D mass profile that agrees simultaneously
with both the lensing and X-ray measurements. However, the projected
temperature profile predicted with this 3D mass profile exceeds the
observed temperature by $\sim 30\%$ at all radii, a level of
discrepancy comparable to the level found for other relaxed
clusters. This result may support recent suggestions from
hydrodynamical simulations that denser, more X-ray luminous
small-scale structure can bias observed temperature measurements
downward at about the same ($\sim 30\%$) level. We determine the gas
entropy at $0.1r_{\rm vir}$ (where $r_{\rm vir}$ is the virial radius)
to be $\sim 800$ keV cm$^2$, as expected for a high temperature
cluster, but its profile at $>0.1r_{\rm vir}$ has a power-law form
with index $\sim 0.8$, considerably shallower than the $\sim 1.1$
index advocated by theoretical studies and simulations.
Moreover, if a constant entropy ``floor'' exists at all, then it is
within a small region in the inner core, $r<0.02r_{\rm vir}$, in
accord with previous theoretical studies of massive clusters.

\end{abstract}


\section{Introduction }
\label{introduction}

As the largest gravitationally bound systems displaying a range of
distinct observational phenomena, clusters of galaxies provide
information of central importance in cosmology. Total gas and dark
matter masses and their profiles provide significant insight into the
formation and evolution of clusters and the relationship between
baryons and dark matter. It has become increasingly clear that much
more can be learned from careful comparisons of cluster observables
including galaxy motions, gas properties from X-ray and
Sunyaev-Zel'dovich (SZ) measurements, and mass profiles from lensing
distortions and multiple images, providing new insight and tighter
constraints on the dynamical state of a cluster and the nature of dark
matter.

Chandra and XMM observatories have provided very detailed information
on the physical state of the ubiquitous X-ray emitting plasma found in
clusters. It has become clear that, broadly speaking, there are two
classes of clusters, those showing some evidence of interaction,
evidenced by complex structures in the gas, and those for which the
gas emission is symmetric with a smooth radial variation in
temperature and abundance, indicating the gas is probably relaxed and
in a hydrostatic equilibrium with the overall gravitational
potential. For lensing work the HST/ACS allows the inner caustic
structure and central mass distribution to be examined in detail
(Gavazzi et al.\ 2002; Broadhurst et al.\ 2005a; Sharon et al.
2005). The wide field imagers such as on the Subaru and CFHT
telescopes permit a statistically significant detection of weak
lensing distortion and magnification effects on the background
galaxies to be traced out to the outskirts of the cluster (Gavazzi et
al.\ 2004; Kneib et al.\ 2005; Broadhurst et al.\ 2005b).

In the case of interacting clusters it has proved very interesting to
compare their lensing-based mass distribution with their disturbed gas
distribution. Clear evidence has emerged in the most favorable case of
1E0657-56 (the "bullet cluster") that two massive clusters have
recently collided in the plane of the sky with a high relative
velocity, leaving the gas lying in-between a bimodal distribution of
galaxies and dark matter, traced by weak lensing (Markevitch et al.\
2002; Clowe, Gonzalez \& Markevitch 2004; Bradac et al.\ 2006). This
particular case demonstrates that the bulk of the mass is dark and
relatively collisionless, as anticipated in CDM dominated cosmogonies
(Markevich et al.\ 2004, Clowe et al.\ 2006, Milosavljevic et al.\
2007, Randall et al.\ 2007), although the estimated relative velocity
between the two massive components may be exceptional in the context
of $\Lambda$CDM simulations (Hayashi \& White 2006). Other such
examples are coming to light, with collisions closer to the line of
sight (Czoske et al.\ 2002, Jee et al.\ 2007, Dupke et al.\ 2007).  A
detailed lensing and X-ray study of a larger sample of interacting
clusters by Okabe \& Umetsu (2007) spans the full range of dynamical
interaction, from premergers where the gas is clearly unaffected by
the mutual gravitational attraction, to cases where both mass
components are still readily distinguishable but the gas heavily
disrupted and shock heated, and finally the postmerger phase where the
gas shows only local signs of interaction, with a relatively small
degree of substructure visible in the dark matter as probed by
lensing.

Strong lensing based masses of the central regions of massive clusters
have often been significantly higher than central masses deduced from
X-ray analysis, by factors of $\sim 2-4$ (Miralda-Escude \& Babul
1995, Wu \& Fang 1997; Allen 1998, Wu et al.\ 1998, Voigt \& Fabian
2006). Obvious reasons for the different masses - in addition to
modeling and intrinsic observational uncertainties - include possible
deviations from hydrostatic equilibrium, sphericity and gas
isothermality (see Allen 1998). Lensing estimates are naturally biased
upward in the case where the gravitational potential is significantly
elliptical with the major axis preferentially aligned towards the
observer. Evaluation of this effect in the context of $\Lambda$CDM
simulations shows that the inferred concentration of the cluster
profile, measured in terms of the Navarro, Frenk \& White (1996,
hereafter NFW) model, can be enhanced by up to 20\% by this bias in
the worst cases (Hannawi et al.\ 2005, Oguri et al.\ 2005), falling
well short of the reported discrepancies.
 
Motivated by the need to improve the precision of measurements of gas
and DM profiles of clusters, we have begun a program of joint X-ray,
strong and weak lensing analyses of several clusters for which we can
combine high quality resolved observations. The advantages of a
simultaneous analysis of X-ray and lensing data are clear, given that
strong lensing measurements yield the total mass profile in the inner
cluster core while X-ray and weak lensing measurements cover a much
larger region of the cluster. The increasing quality and degree of
detail of such data allows a more model-independent determination of
the relevant profiles along with redundancy so that self-consistency
can be checked. Under the assumption of spherical symmetry and
hydrostatic equilibrium the projected temperature and gas density
profiles as well as the total surface mass density profile may now be
derived directly without resorting to assumed models or simple
parameterizations of the profiles.

Here we apply a model independent approach to derive the density
profile of a relaxed cluster from a simultaneous fit to both the X-ray
and lensing data.  We apply our technique to A1689 ($r_{vir} \sim 2$
h$^{-1}$ Mpc), a rich and moderately-distant ($z=0.183$) cluster that
has been extensively observed in the optical, near IR and X-ray
regions, and is the first cluster in our sample.  The cluster has a cD
galaxy whose center is within $\sim 1.5"$ of the X-ray centroid; this
fact, and the low degree of X-ray ellipticity ($\epsilon\simeq 0.08$,
Xue \& Wu 2002, hereafter XW02) indicate that the cluster is likely to
be well relaxed and nearly spherical.  Previous estimates of its mass
were obtained from the analysis of observed arcs and arclets produced
by strong lensing (Broadhurst et al.\ 2005a), from the distortion of
the background galaxy luminosity function and number density (Taylor
et al.\ 1998, Dye et al.\ 2001), and from weak lensing observations
(Broadhurst et al.\ 2005a,b, Medezinski et al.\ 2007).  Here we derive
the X-ray surface brightness and temperature profiles from an analysis
of the full set of Chandra observations, which can be compared with a
smaller subset of Chandra observations analyzed by XW02, and an
independent study based on XMM (Andersson \& Madejski 2004; hereafter
AM04).

The X-ray and lensing observations and data reduction are described in
Section~2, followed by a detailed account of the spectral and spatial
data analysis in Section~3. In Section~4 we describe the methodology
of deriving the gas and mass profiles, and in Section~5 we present the
results of our deduced gas, total mass, and entropy profiles. Our
results are discussed and assessed in Section~6.

\section{Observations and Data Reduction}
\subsection{X-ray measurements}
\label{sec:X}

A1689 was observed by Chandra during three non-consecutive periods. 
These observations were made mainly with the onboard Advanced
CCD Imaging Spectrometer in $2\times 2$ imaging array (ACIS-I)
mode. Table \ref{Observation time data} gives a summary of the data 
we have analyzed including the Good Time Interval 
(GTI, i.e., exposure time after all known corrections were applied). 
We reduced all data using the following release of data reduction 
software: Chandra data analysis software package CIAO 3.3, with the 
updated complement calibration database CALDB 3.2. 
\footnote{We followed the threads for data preparation "Analysis Guide: 
ACIS Data Preparation"
http://cxc.harvard.edu/ciao/guides/acis\_data.html, and for extended
sources "Analysis Guide: Extended Sources",
http://cxc.harvard.edu/ciao/guides/esa.html. Due to the risk that some
cluster emission extends over the entire image we took the background
from the ACIS "Blank-Sky" Background files compiled by Markevitch
(2001), and followed the thread "Using the ACIS "Blank-Sky" Background
Files", http://cxc.harvard.edu/ciao/threads/acisbackground/, and
Maxim's cookbook, http://cxc.harvard.edu/cal/Acis/Cal\_prods/
bkgrnd/acisbg/COOKBOOK.} All observations were reduced from the level
1 stage in order to achieve a better modeling of instrument gain and
quantum efficiency. The event grades which we used are GRADE=0, 2, 3,
4 and 6. Periods of background flaring were removed using the CIAO
task "lc\_clean". We removed bright sources using the tool "wavdetec"
with the default parameters: scales="2.0 4.0" and sigthresh =
$10^{-6}$. Results of setting scales="1.0 2.0 4.0 8.0 16.0" were
checked for Observation ID 540; no other bright sources were
identified. We tailored the background files to the corresponding data
sets. Identical spatial filters were applied to the sources and the
background data sets. Spectra and responses were extracted using the
tool "specextract". All channel count rates were combined into one
spectrum (using the FTOOL task MATHPHA, and matching response matrix
and ancillary response files, using the tools ADDRMF and ADDARF,
respectively), weighting individual exposures by their respective
integration times. We then binned the combined counts so that there
were at least 25 counts per bin. The center of the cluster was found
by IRAF to be at $13^h11^m29.^s575-
01^\circ20^\prime27.^{\prime\prime}59$, in agreement with the position
determined by AM04 ($13^h11^m29.^s4-
01^\circ20^\prime28^{\prime\prime}$).

\begin{table*}

\caption{{\it Chandra} Observation Log for A1689 \label{Observation time data}}
\begin{center}
\begin{tabular}{|c|c|c|c|c|}

\hline
 Obs. ID            & Start time                       & Mode & Duration & Good time interval (GTI) \\
                    &                                  &      &  [ks]        &   [ks] ACIS    \\
\hline                    
 540                & 2000-04-15 04:13:33              &FAINT&  10.45&  10.32   \\
 1663               & 2001-01-07 08:18:34              &FAINT&  10.87&  10.62   \\
 5004               & 2004-02-28 07:18:29              &VFAINT&  20.12& 18.44   \\ 
                    &                                  &     &  41.44 &  39.38  \\
\hline
\end{tabular}
\end{center}
\end{table*}

The spectrum of each observation was first checked in order to
identify possible problematic features, and to assess the consistency
between the model parameters obtained from the different observations.
We reduced the spectrum within a circular radius of $3^\prime$
(corresponding to a physical radius of $387 h^{-1}$ kpc), and binned
each spectrum to have at least 20 counts per bin. Using XSPEC we
fitted the $0.3-10$ keV data to an optically thin thermal plasma model 
with Galactic photoelectric absorption, WABS(MEKAL), with $N_{H}= 
2\cdot10^{20}$ cm$^{-2}$, the mean absorption along the line of sight 
to A1689 (Dickey \& Lockman 1990). The resulting parameter values were
consistent between the fits, but the fit quality varied, with
$\chi^{2}_{r}=$ 1.231, 1.366, and 1.478 for observation ID 540, 1663,
and 5004, respectively. Deviations between the data and model were
high in two energy bands. Indeed, in the $7-10$ keV band the count
rate is significantly higher than the model prediction, as noted also
by AM04 (and as also seen in fig.~2 of XW02). This could partly 
be due to a background of high energy particles, but filtering the
background of observation ID 540 using a smaller time binning did not
lower the difference in the count rate. Uncorrected instrumental
effects can also be invoked, such as an imprecise correction for the
contaminating lines from the external calibration source\footnote{For
more on this, see http://cxc.harvard.edu/cal/Acis/
Cal\_prods/bkgrnd/current/}.

The second problematic band was $0.3-0.5$ keV, where the data in all
three observations are higher than the model values (even if
absorption is ignored). As mentioned in AM04, there is extra
absorption caused by molecular contamination of the ACIS optical
blocking filters which causes the data to be lower than the model. A
correction is implemented in the analysis software (starting with
version CIAO 3.0), but because of the large uncertainty in the ACIS
gain at energies below 350 eV, the recommended procedure is to ignore
events in the 0.3-0.35 keV band\footnote{http://www.astro.psu.
edu/users/chartas/xcontdir/xcont.html}. We do not know if a
substantial uncertainty extends also to $0.35-0.5$ keV, so to be safe,
we ignored the $0.3-0.5$ keV data, and used only the $0.5-7$ keV
measurements.

The change in energy interval for the ID 5004 observation resulted in
only $<1\%$ difference in values of the fitted parameters. This "band
stability" increases the confidence in the reliability of the results.
However, the temperature is somewhat sensitive to the specific value
used for the Galactic absorption. Fits with $N_{H}$ as a free
parameter yield a reasonable value, $1.32\cdot 10^{20}$ cm$^{-2}$ for
observation 540, but a value which is close to zero for the two other
observations. Also, $kT$ is higher by $\sim 1$ keV than in the fit
with $N_{H}$ fixed at its observed value ($2\cdot 10^{20}$
cm$^{-2}$). This is further discussed in section~6 below.  The
measured flux and surface brightness are insensitive to this change,
since the overall normalization does not change when the absorption is
taken to be a free parameter.

More Chandra observations of A1689 have recently become public, namely
observations ID 6930 and 7289, totaling 80 ks. We have checked the
quality of these new data by deriving the temperature profile using
CIAO 3.4. The deduced value is systematically offset higher by $1.5-2$
keV than the value obtained from the earlier observations, and from
values obtained from measurements with other instruments. Also, the
profile from the new observations does not show a decrease at large
radii but rather an increase. The problem is even more severe using
CIAO 3.3 and is due to improper background files. These results
indicate that there still is a problem with the matched background
files. We have therefore not included these observations in our work.

\subsection{Lensing measurements}

Analysis of strong lensing measurements of A1689 was carried out using
deep HST/ACS images with a total of 20 orbits shared between the GRIZ
passbands. Over 100 lensed background images have been identified,
corresponding to 30 multiply-imaged background galaxies, including
many radial arcs and small de-magnified images inside the radial
critical curve, close to the center of mass (Broadhurst et al.\
2005a). For a given lens model Broadhurst et al.\ projected the lensed
images onto a sequence of source planes at various distances. They
then generated model images by lensing these source planes and
comparing the detailed internal structures of observed images falling
near the predicted model positions, where the unknown source distance
is a free parameter for each source. As new images were identified
they were incorporated into the lens model to refine it, enhancing the
prospects of finding additional lensed images. This relatively rich
lensing field allows a good mapping of the mass in the inner core.

At larger radius the statistical effects of weak lensing have been
used to explore the entire mass profile of A1689 using wide-field V
and I-band images taken with Subaru/Suprime-cam (Broadhurst et al.\
2005b). In practice this work is difficult, requiring careful analysis
of large sets of wide field images, with corrections for seeing,
tracking and instrumental distortion. Broadhurst et al.\ obtained a
firmer constraint on the mass profile by examining the effect of
lensing on the background number counts of red galaxies, as advocated
by Broadhurst, Taylor \& Peacock (1995). Broadhurst et al.\ found a
clear detection of both the weak distortions and a deficit in the
number counts to the limit of the Subaru images, which corresponds to
an outer radius of $r\sim 1.5$ h$^{-1}$ Mpc (Broadhurst et al.\ 2005b).

Combining the inner mass profile derived from strong lensing with the
outer mass profile from weak lensing we see that for A1689, the
projected mass profile continuously flattens towards the center like
an NFW profile, but with a surprisingly steep outer profile
(Broadhurst et al.\ 2005b, Medezinski et al.\ 2007) compared with the
much more diffuse, low concentration halos predicted for massive CDM
dominated halos, (e.g., Bullock et al.\ 2002).

\section{Spectral and Spatial Data Analysis}

The X-ray flux was measured in two bands - $0.5-9$ keV, which was used
to check overall consistency with previous observations, and the
narrower band $0.5-7$ keV used in this work (see section~2.1).  In a
$3^\prime$ aperture the measured flux in the latter band was
$F_{0.5-7keV}=(2.17\pm0.01)\cdot10^{-11}$ erg cm$^{-2}$ s$^{-1}$,
slightly lower than in the wider band, $F_{0.5-9keV}=
(2.44\pm0.01)\cdot10^{-11}$ erg cm$^{-2}$ s$^{-1}$. The
corresponding luminosities were $L_{0.5-7keV}=(1.002\pm0.004)\cdot
10^{45}$ h$^{-2}$ erg s$^{-1}$, and
$L_{0.5-9keV}=(1.128\pm0.004)\cdot 10^{45}$ h$^{-2}$ erg s$^{-1}$.
These values are consistent with previous findings. XW02, who analyzed
some of the Chandra data, obtained
$F_{0.5-10keV}=(2.7\pm1)\cdot10^{-11}$ erg cm$^{-2}$ s$^{-1}$ and
$L_{0.5-10keV}=(1.03\pm0.38)\cdot 10^{45}$ h$^{-2}$ erg s$^{-1}$.
Our flux is also consistent with that measured by XMM (AM04) and ROSAT
(Ebeling et al.\ 1996).

The degree of ellipticity of the cluster X-ray emission is not only a
good indication of its 3D morphology, but also gives some indication
of its dynamical state. These important properties are of particular
relevance to our work which is based on the assumptions that the gas
is roughly spherically symmetric and that the cluster is in a state of
hydrostatic equilibrium. Using the SExtractor utility we estimated the
ellipticity of the X-ray emission to be $\epsilon= 0.083\pm 0.002$,
which is in very good agreement with the value reported by XW02.

\begin{figure}
\centerline{\includegraphics[width=8.0cm]{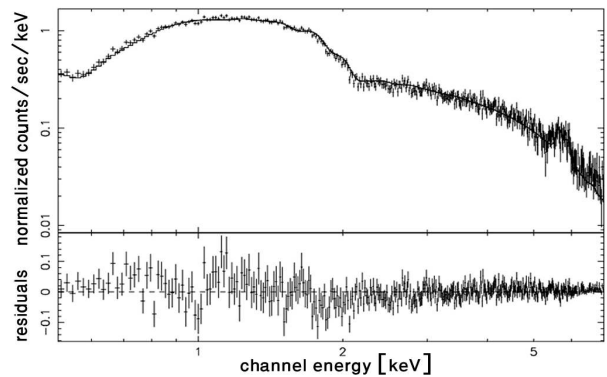}}
\caption{We show (upper panel) the observed X-ray spectrum of A1689 from the 
combined observations ID 540, 1663 and 5004 in the $0.5-7$ keV band,
reduced from a $3^\prime$ aperture. The fit (solid curve) is to a
MEKAL model with Galactic absorption fixed at the observed
value. We also show (lower panel) the residuals of the fit. \label{540and1663and5004_0.5_7kev_3arcmin fig}}
\end{figure}

We measured the cluster gas temperature from the spectral measurements
(Figure~\ref{540and1663and5004_0.5_7kev_3arcmin fig}). To check
consistency of the mean temperature $T$ with previously determined
values, we analyzed the combined dataset in the $0.3-9$ keV band
(which is, as noted above, wider than the more uniform $0.5-7$ keV
dataset used in the rest of our work). The fit (with
$\chi^{2}_{r}=1.14$) yielded $kT=9.36\pm 0.18$ keV, and a metal
abundance $A=0.4\pm 0.04$ in solar units, consistent with
$kT=9.02^{+0.4}_{-0.3}$ keV (Mushotzky \& Scharf 1997), and
$kT=8.2-10$ keV, $A=0.2-0.49$ (XW02). In the $0.5-7$ keV band, the
corresponding values are $kT=9.35\pm 0.18$ keV, and $A=0.41\pm 0.04$,
and $\chi^{2}_{r}=1.03$. If Galactic absorption is treated as a free
parameter, an unrealistically low value is determined, and $kT=
10.66_{-0.37}^{+0.42}$ keV. In
table~\ref{temperature_table_and_Galactic_absorption} we list $kT$ for
each observation and the value from the combined dataset with Galactic
absorption either fixed at the observed value, or treated as a free
parameter.

In the following subsections we present the X-ray derived profiles of
the metal abundance, surface brightness, and temperature.

\begin{table*}

\caption{Mean gas temperature over a $3^\prime$ region for each 
observation, with Galactic absorption either fixed at the 
observed value (Dickey \& Lockman 1990) or left as a free parameter.
 \label{temperature_table_and_Galactic_absorption}}. 
\begin{center}
\begin{tabular}{|c|c|c|c|c|}

\hline
        & $N_{G}=2\cdot 10^{20}$ cm$^{-2}$ & & $N_{G}$ is a free parameter &\\
 Obs. ID&$\chi_{r}^{2}$&   T [keV]           &  $\chi_{r}^{2}$  & T [keV] \\              
\hline                    
540     &  1.16  & 9.7$_{-0.5}^{+0.8}$ & 1.16 & 10$_{-0.4}^{+0.7}$     \\
1663    &  1.1   & 9$_{-0.5}^{+0.5}$   & 1.01 & 10.2$_{-0.6}^{+0.4}$   \\
5004    &  1.19  & 9.3$_{-0.4}^{+0.5}$ & 1.13 & 10.5$_{-0.4}^{0.3}$   \\
combined&  1.03  & 9.4$_{-0.3}^{+0.4}$ & 0.95 & 10.7$_{-0.2}^{+0.3}$   \\
\hline
\end{tabular}
\end{center}
\end{table*}

\subsection {Heavy element abundances}
\begin{figure}
\centerline{\includegraphics[width=8.5cm]{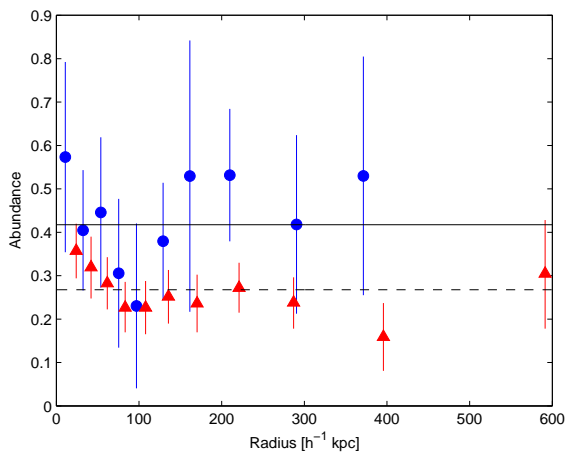}}
\caption{The abundance profile as derived from the spectral deprojection 
fitting. We compare our Chandra analysis (circles), which gives a mean
value of $0.42$ (horizontal solid line), to the XMM analysis of AM04
(triangles) which gave a mean value of $0.27$ (horizontal dashed line).
\label{Abundance gradient}}
\end{figure}

Before assessing the observed temperature profile it is important to
determine the gas metal abundance.  In principle, these two quantities
may be decoupled with high quality data; however, the limited spectral
resolution and signal-to-noise ratio means that the abundance is
generally hard to constrain independently with radius, and one must
adopt a mean value for the cluster. Vikhlinin et al.\ (2005) took
advantage of the superior spatial resolution of Chandra to show that
for a sample of nearby relaxed clusters there is a metallicity
gradient such that the abundance increases toward the center. However,
the high roughly solar abundance level is actually in a region
coinciding with the central galaxy. On the theoretical side, Arieli,
Rephaeli \& Norman (in preparation) have performed a high resolution
hydrodynamical simulation which uses a new approach to incorporate
feedback from galaxies on the intracluster gas, and have shown that
including physical processes such as galactic winds and gas stripping
yields a flat metallicity profile out to large radii ($\sim 600$ h$_{
0.7}^{-1}$ kpc). Based on XMM data, AM04 showed that there is no
abundance gradient in A1689. Our deduced abundance gradient is shown
in figure~\ref{Abundance gradient}. There is a systematic trend
towards a higher value of $A$ in the Chandra data (in accord with the
result of XW02) than the corresponding XMM value, but this difference 
is not very significant from a statistical point of view, since most 
of the data points agree within the $1\sigma$ range. The two sets of 
observations have the same exposure time, so the XMM data are more 
precise (due to the larger effective area).  Our analysis indicates 
a fairly constant abundance over the region probed. In what follows 
we use the mean value of $0.4$ solar for the central $3^\prime$ region.

\subsection{Surface brightness and temperature analysis}
\begin{figure}
\centering
\epsfig{file=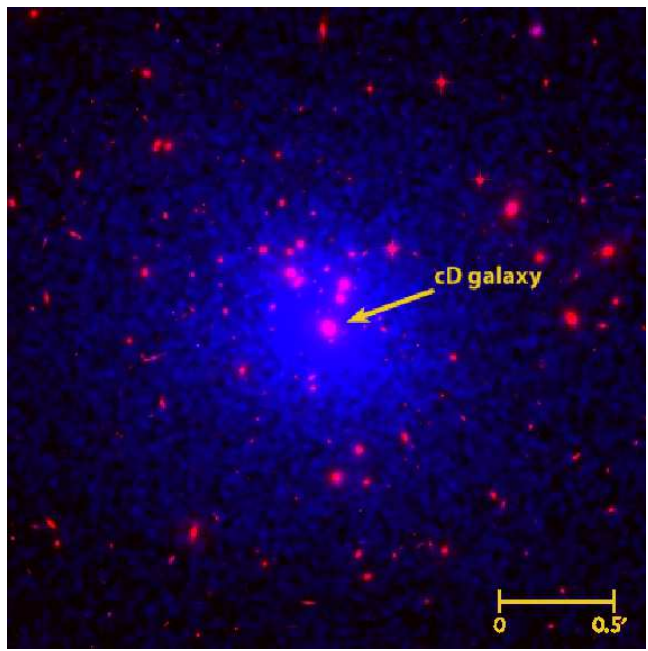, width=8.5cm}
\caption{HST/ACS image of A1689 overlaid on smoothed color rendering of the 
X-ray emission (obs. ID 5004). The cD galaxy is seen to coincide with the 
X-ray centroid. The scale in the image is $0.5^{\prime}\simeq 65$ h$^{-1}$ 
kpc.
\label{xray_and_optics}} 
\end{figure}

An HST/ACS image of A1689 overlaid on the X-ray map is shown in
figure~\ref{xray_and_optics}; the cD galaxy lies at the X-ray
centroid.

The symmetry of the surface brightness distribution allowed us to
construct an azimuthally averaged profile in radial bins. We
determined an azimuthally averaged flux $F$ in each radial bin by
fitting the model to the data in each annulus and extracting the flux,
from which the surface brightness $S$ was found by dividing by the
angular area of the bin. The error in $S$ was calculated from that in
the normalization parameter of the fitted model. Note that our profile
of the surface brightness is $\sim 4$ times higher than that obtained
by XW02 (see their Figure~6), but is consistent with that obtained by
Mohr, Mathiesen \& Evrard (1999, hereafter MME99; see their Figure~13).

The full extraction region consists of 50 annuli, each of width
4$^{\arcsec}$, plus nine annuli with varying widths (to keep adequate
S/N) beyond 200$^{\arcsec}$, all centered on the cluster X-ray center
(section~2.1). The maximum aperture of the data reduction region was
limited by the outer edges of CCDs in practically all three
observations. We removed the innermost ring because it corresponds to
the inner region of the cD galaxy (see also Vikhlinin et al.\ (2005);
e.g., the abundance in it was measured to be $0.9$, contrary to our
assumption of a constant abundance in each annulus.

The reduction of the temperature profile was done in a similar way,
but with a smaller number of annuli due to the lower signal-to-noise
ratio.

\section{Derivation of the Gas and Mass Profiles}
\subsection{Methodology}

In this section we present our procedure for studying the structure of
the cluster using all the available data sets in combination. Since we
have both lensing and surface brightness profiles, we do not need to
assume any particular parametrization or fitting formula, such as the
commonly adopted NFW mass profile or the $\beta$ model for the gas
density profile. We employ a model-independent approach that is
limited only by the resolution and accuracy of the data. As mentioned
in the introduction, the cluster has only a small ellipticity, so the
assumption of spherical symmetry is reasonable. This basic assumption
allows us to reconstruct three-dimensional profiles from the observed
two-dimensional ones. The lensing data yields in this way the density
profile of the total mass (dark matter plus gas), while the surface
brightness profile depends on a combination of the gas density and
temperature. Thus, with one additional relation, we can reconstruct
the full gas and dark matter density profiles. We derive this
additional relation by assuming hydrostatic equilibrium, consistent
with the indications discussed in section~1 that A1689 is well
relaxed. Hydrostatic equilibrium involves the gravitational force and
thus inserts a dependence on the total mass profile that couples the
constraints from lensing and from the X-ray data.

We determined the best-fit values of our free parameters by fitting
the lensing and X-ray surface brightness data simultaneously. It is
important to note that we did not include the temperature profile data
for reasons that are explained later in this section. In our
model-independent approach, our free parameters are the values of the
3D profile of the total mass density and the gas mass density at
several fixed radii (logarithmically spaced). The radial ranges of the
free parameters of the total mass density and the gas density were
determined by the range of the lensing data and the X-ray surface
brightness data, respectively. Within these ranges the density values
were linearly interpolated (in log-log) in-between the fixed
radii. Beyond the last data point, each profile (i.e., of the total
mass density and the gas density) was extrapolated. The total mass
density profile was extrapolated as $\rho(r) \propto r^{-3}$, in
accordance with the asymptotic behavior of the NFW profile and also
close to our best-fit core plus power-law model (section~5.2
below). The gas density was also extrapolated as $\rho(r) \propto
r^{-3}$, in order to have a constant $f_{\rm gas}$ at large radii;
this is also consistent with the previously deduced index of 2.9 (Xue
\& Wu 2000). 

In spherical symmetry, the hydrostatic equilibrium equation for an
ideal gas can be integrated from a given three-dimensional radius $r$
out to infinity (or some maximum, cutoff radius). This yields the
relation
\begin{equation} \left[ \rho_{\rm gas}(r)T(r)\right] \Big|^{r}_{\infty} =
\int^{\infty}_{r}\frac{GM_{\rm tot}(\leq r')\mu m_{p} 
\rho_{\rm gas} (r')} {kr'^2}\,dr'\ , 
\label{eq: HEE}\end{equation} where $\rho_{\rm gas}$ is 
the gas mass density, $T$ is the gas temperature, $M_{\rm tot}(\leq
r)$ is the total mass within $r$, $\mu$ is the mean molecular weight,
and $m_{\rm p}$ is the proton mass. Of course, $\mu$ depends on the
abundances; the typical metal abundance is $\approx\frac{1}{3}$ (which
is expected when ejecta are from Type Ia supernovae winds (Silk
2003)). In A1689, $A=0.2-0.49$ was found by XW02.

For given model parameters, the comparison to the lensing data was
performed by first projecting the 3D profile of the total mass density
using Abel integration
\begin{equation} \kappa(R)  = \frac{2}{\Sigma_{crit}} \int^{\infty}_{R} 
\frac{\rho_{\rm tot}(r) r dr}{\sqrt{r^{2}-R^{2}}}\ ,\label{eq: kappa}
\end{equation} where $\rho_{\rm tot}(r)$ is the three dimensional total mass 
density and $\Sigma_{crit}=\frac{c^2}{4\pi G}\frac{D_{\rm os}}{D_{\rm
ol} D_{\rm ls}}$ is the critical density for lensing, written in terms
of angular diameter distances $D_{\rm os}$ (observer--source), $D_{\rm
ol}$ (observer--lens), and $D_{\rm ls}$
(lens--source). Simultaneously, we calculated the mass profile using
\begin{equation} M_{\rm tot}(\leq r) =
  4\pi \int^{r}_{0} \rho_{\rm tot}(r^\prime) r^{\prime 2} dr^\prime\ .
\label{eq: M}  \end{equation}
We then used the mass profile and the gas density profile in the
hydrostatic equilibrium equation (\ref{eq: HEE}), obtaining the
temperature profile. The temperature and gas density profiles were
then used (with the assumed abundances) to determine the emissivity:
\begin{equation}
\varepsilon(r)= n_{e}(r)n_{H}(r)
\Lambda(T(r))\ , 
\end{equation} 
where $\Lambda(T)$ is the cooling function, which was obtained by
considering all the relevant physical processes in the $9\cdot
10^{5}-3\cdot 10^{8}$ K range.  The cooling function was calculated by
MEKAL (consistent with our fitting of the observed spectra in
section~\ref{sec:X}).
We use the usual definitions, $n = \rho_{\rm gas}/(\mu m_{p})$,
$n_{e}=\rho_{\rm gas} /(\mu_{e}m_{p})$, and $n_{H}=\rho_{\rm gas}/
(\mu_{H}m_{p})$, which for $A=0.4$ solar yield $\mu=0.55$, 
$\mu_{e}=1.05$, and $\mu_{H}=1.30$.

Having thus obtained the emissivity, we re-projected it using the
Abel integral and obtained the X-ray surface brightness
\begin{equation} S(R)  = 1.191\cdot 10^{-12}\frac{\pi }{(1+z)^4} 
\int^{\infty}_{R} \frac{\varepsilon(r) r dr}{\sqrt{r^{2}-R^{2}}}\ ,
\label{eq: XSB} \end{equation}
where $\varepsilon$ is the emissivity in erg s$^{-1}$ cm$^{-3}$, $S$
is the X-ray surface brightness in erg s$^{-1}$ cm$^{-2}$
arcsec$^{-2}$, and we converted the units from erg s$^{-1}$
cm$^{-2}$ to erg s$^{-1}$ cm$^{-2}$ arcsec$^{-2}$ by multiplying by
$\frac{D_{A}^2}{4\pi D_{L}^2}$, using $D_{L}/D_{A}=(1+z)^2 $ (where
$D_{L}$ and $D_{A}$ are the angular diameter and luminosity distance,
respectively), and upon conversion of radians$^{-2}$ to arcsec$^{-2}$.
The mean surface brightness was calculated 
in each annulus-shaped bin within the surface brightness data. The mean 
surface brightness $S_{\rm bin}$ in a bin with inner and outer radii 
$R_{\rm in}$ and $R_{\rm out}$, respectively, is
\begin{equation} 
S_{\rm bin} = \frac{2}{(R_{out}^{2}-R_{in}^{2})} 
\int^{R_{out}}_{R_{in}} R S(R)dR\ . \label{eq:XSB2} 
\end{equation} 
Finally, we compared this to the surface brightness data.

Having calculated the observed quantities for each possible set of
model parameters, we then compared to the data using a $\chi^2$
measure. The total $\chi^2$ is the sum of two terms, one from the
lensing data and one from the X-ray surface brightness data:
\begin{equation} 
\chi^{2} =
  \chi^{2}_{\kappa}+\chi^{2}_{S}\ ,\label{eq: chi^{2}}\end{equation} 
  where
  \begin{equation} \chi^{2}_{\kappa} = \sum_{i=1}^{N_{\kappa}} \left[
\frac{\kappa(R_{i})-\kappa_{data}(R_{i})}{\Delta
\kappa_{data}(R_{i})} \right]^{2}\end{equation} and
\begin{equation} \chi^{2}_{S} = \sum_{j=1}^{N_{S}} \left[
\frac{S_{\rm bin}(R_{j})-S_{data}(R_{j})}{\Delta S_{data}(R_{j})}
\right]^{2}\ .\end{equation} Here $\chi_{\kappa}^{2}$ is the $\chi^2$
of the lensing data, obtained by comparing the model $\kappa$ of
equation~(\ref{eq: kappa}) with the $N_{\kappa}$ data points
$\kappa_{data}$ and their errors $\Delta \kappa_{data}$; similarly,
$\chi_{S}^{2}$ is the $\chi^2$ of the X-ray surface brightness data,
obtained by comparing the model $S$ of equation~(\ref{eq:XSB2}) with
the $N_{S}$ data points $S_{data}$ and their errors $\Delta S_{data}$.
The best-fit values of the parameters, which are the 3D profile of the
total mass density and the 3D gas density profile, were obtained by
minimizing the total $\chi^2$. From these two profiles we then derived
the three dimensional temperature profile using equations~(\ref{eq:
HEE}) and (\ref{eq: M}). In order to perform an independent
consistency check with the two-dimensional temperature data (which
were not used in the analysis), we projected the 3D temperature,
weighting it by the emissivity for comparison with the observed
temperatures:
\begin{equation} T_{\rm 2D}(R) =
\frac{\int^{\infty}_{R} T(r)\, \varepsilon(r)\, \frac{r}
{\sqrt{r^{2}-R^{2}}}\, dr}{\int^{\infty}_{R} \varepsilon(r)\,
\frac{r}{\sqrt{r^{2}-R^{2}}}\, dr }\ . 
\end{equation} 
Weighting by the emission measure EM (which is proportional to
$\rho_{\rm gas}^2$) gave similar results.

The EM weighted temperature ($T_{EM}$) and the emissivity weighted
temperature ($T_{E}$) may not precisely reflect the actual spectral
properties of the observed source. Vikhlinin (2005) states that the
"spectroscopic" temperature is generally lower than either of these
temperatures, although his suggested correction formula for the
effective spectroscopic temperature of a multi-component thermal
plasma yielded similar values to $T_{EM}$ and $T_{E}$. Mathiesen \&
Evrard (2001) claimed that the observed spectroscopic temperature may
be biased low due to the excess of soft X-ray emission from the small
clumps of cool gas that continuously merge into the intracluster
medium. Furthermore, Mazzotta et al.\ (2004) showed that from a purely
analytical point of view the spectrum of a multi-temperature thermal
model cannot be accurately reproduced by any single-temperature
thermal model. It has also been found in simulations that the spectral
temperature is in general smaller than $T_{E}$ by a factor of
$0.7-0.8$ (Rasia et al.\ 2005). We note that total mass estimates
obtained based solely on X-ray data together with the assumption of
hydrostatic equilibrium depend strongly on the temperature. In our
analysis, however, we do not use the temperature data for the fit but
only for a consistency check. Therefore, our values for the total mass
density are determined mainly by the lensing data.

Our analysis focuses on the free-parameter method that does not assume
a particular shape for the profile.  In addition, we used our analysis
method to test the viability of either an NFW profile or a cored
profile.

\subsection{The entropy profile}

It is interesting to evaluate the entropy and the adiabatic index
profiles in order to probe the processes that govern the thermal state
of intracluster gas.  For a monoatomic ideal gas $s=\frac{3}{2}k \ln K
+s_{0}$, where $s=\frac{S}{N}$ is the entropy per particle and
$K\equiv T\, n^{-2/3}$. In this case $K$ is a combination of $n$ and
$T$ that is invariant under adiabatic processes in the gas.  More
generally, $T\propto n^{\Gamma-1}$ under adiabatic changes, where
$\Gamma$ is the adiabatic index. In this paper we refer to $K$ as the
``entropy''.

The gas at $\sim 0.1R_{\rm vir}$ lies outside of the cooling region,
and is still sufficiently close to the cluster center where shock
heating is minimized (Lloyd-Davis, Ponman, \& Cannon 2000). Thus at
this radius, it is less sensitive to specific models and
assumptions. In the outer regions, $r\gtrsim 0.1R_{\rm vir}$, the
entropy is entirely due to gravitational processes, and the entropy
profile is expected to be a featureless power law approaching $K
\propto r^{1.1}$ (Tozzi \& Norman 2001; Voit, Kay, \& Bryan 2005). In
the inner region the gas entropy profile is flattened (for ASCA and
ROSAT data, see Lloyd-Davis, Ponman, \& Cannon 2000; Ponman,
Sanderson,\& Finoguenov 2003; for various cosmological simulations see
Voit, Kay, \& Bryan 2005; for a theoretical analysis of joint X-ray
and SZ observations, see Cavaliere, Lapi, \& Rephaeli 2005) .

\subsection{The effect of the cD galaxy}
\label{sec: cD}

A massive cD galaxy dominates the inner region of A1689.  Thus, it is
important to check the effect of the cD galaxy on the cluster, and
especially on the 2D temperature observed in the central region. Its
mass profile can be analytically expressed as follows. The mass
profile is $M_{\rm cD}=M_{\rm stars}+M_{\rm DM}$, where the gas mass
is neglected since it is dominated by the stellar mass at small radii
and by the DM mass at large radii. We assume an NFW profile for the
dark matter, and express the stellar mass profile as $M_{\rm stars} =
M_{\rm stars,0}\cdot \frac{r^2}{(r+a)^2}$, where $M_{\rm stars,0}$ is
the mass of the stars in the cD galaxy and $a$ is a characteristic
length-scale (Hernquist 1990). Typical values for the effective
radius, $R_e$, with $a = R_{e}/1.82$, are $R_e \approx 3-5$ h$^{-1}$
kpc. E.g., Jannuzi, Yanny, \& Impey (1997) found an average $R_e=4.6$
h$^{-1}$ kpc for five elliptic galaxies; Brown \& Bregman (2001) found
an average $R_e=3.2$ h$^{-1}$ kpc for four elliptical galaxies (but
see Arnalte Mur, Ellis, \& Colless (2006) who found $R_e=24.7$
h$^{-1}$ kpc). Typical values for the characteristic radius of the NFW
profile of the DM (see also section~5.2 below) are $5-25$ h$^{-1}$ kpc
(Romanowsky et al.\ 2003).

X-ray emission from the cD galaxy may be produced by the stars and
stellar remnants, or by the hot gas. For the stellar component, the
spectrum (dominated by X-ray binaries) is expected to be hard like
that of the cluster ($T_X\sim 10$ keV) and have a minor effect on the
measured 2D temperature, so we did not include it in our model. We
considered the gas emission, which should give a soft spectrum
($T_X\sim 1$ keV) (Brown \& Bregman 2001). We adopted a double beta
model for the gas density profile, as Matsushita et al.\ (2002) fitted
to M87, where we redid the fit to the M87 data, allowing the
parameters to vary freely except that we fixed the $\beta$ of the
extended component to be $0.47$ (based on data from the ROSAT all sky
survey: Bohringer et al.\ 1994). For a given gas mass density and
total mass profile of the cD, we derived the temperature using
equation~(\ref{eq: HEE}). From the gas density and the temperature we
obtained the emissivity using the MEKAL code, assuming a solar metal
abundance.

\section{Results}
\label{results}

\begin{figure}
\centering
\epsfig{file=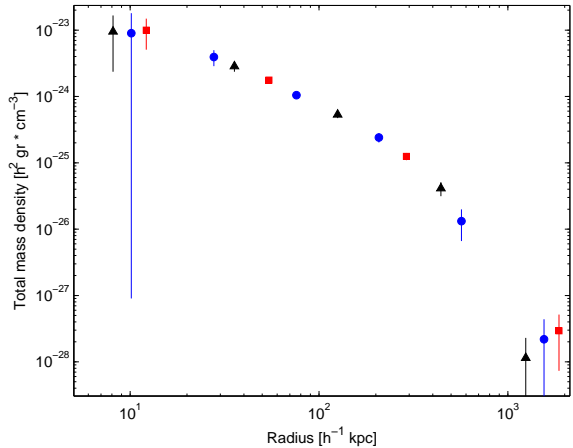, width=8.5cm, clip=}
 \caption{Results for the total mass density profile when different
 numbers of free parameters are used for the profiles of total mass
 ($i$) and gas ($j$). We consider $i=6$, $j=14$ (circles), $i=5$,
 $j=11$ (triangles), and $i=4$, $j=9$ (squares). All three cases had
 the same, fixed values for the smallest and largest radius; these
 sets of points have been offset for display purposes, the triangles
 to the left and the squares to the right.
\label{total_mass_density_different_parameters_sets}}
\end{figure}

\begin{figure}
\centering
\epsfig{file=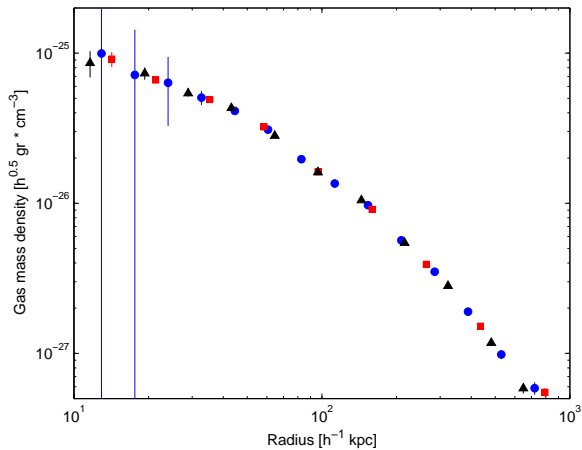, width=8.5cm, clip=}
 \caption{Results for the gas mass density profile when different
 numbers of free parameters are used. Sets of points are as in
 figure~\ref{total_mass_density_different_parameters_sets}. The first
 and last points are offset as in
 figure~\ref{total_mass_density_different_parameters_sets}.
 \label{gas_mass_density_different_parameters_sets}}
\end{figure}

\begin{figure}
\centering
\epsfig{file=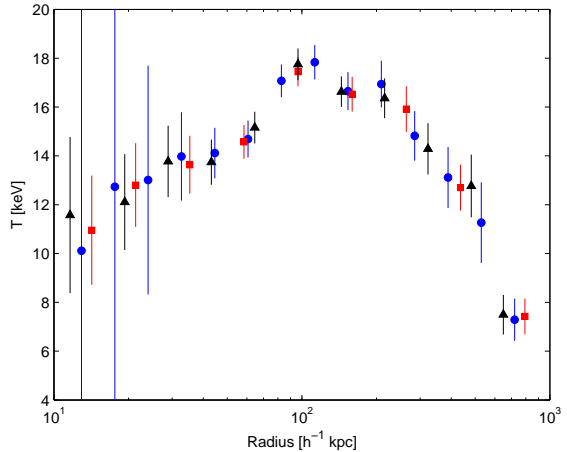, width=8.5cm, clip=}
 \caption{Results for the 3D temperature profile when different
 numbers of free parameters are used. Sets of points are as in
 figure~\ref{total_mass_density_different_parameters_sets}. The first
 and last 
 set of 
 points are again offset as in
 figure~\ref{total_mass_density_different_parameters_sets}.
 \label{temperature_different_parameters_sets}} 
\end{figure}

In this section we present the results of our model-independent
analysis of the lensing and X-ray data. The numbers of data points 
used in the analysis were $N_{\kappa}=26$ and $N_{s}=58$ for lensing
and surface brightness, respectively.  Accordingly, the ratio of the
number of free parameters for the total mass to that for the gas mass
profile was set to $i/j\approx26/58$. As our standard case, we chose 6
free parameters for the total mass density and 14 free parameters for
the gas mass density.  These relatively small numbers of model points
ensured fairly smooth profiles that effectively average over the noise
in the data. The fit to the lensing and surface brightness data can be
seen in fig.~\ref{lensing fit} and fig.~\ref{surface brightness fit},
respectively. Both of the fits are very good, as expressed in the low
reduced $\chi^2$, i.e., $\chi^2_{\rm r} \equiv \chi^2/$dof. The
reduced $\chi^2$ of the fit to the data was 28.1/64.  The contribution
of each data set within the total, simultaneous fit to both, was
$\chi^2_{r}({\rm lensing})=4.7/20$ and $\chi^2_{r}({\rm
surface\;brightness})=23.5/44$. This shows that we achieved a good fit
to both data sets, and thus did not need a larger number of model
points. The low $\chi^2_{\rm r}$, especially in the lensing case,
suggests errors that are either overestimated or significantly
correlated among the various data points, producing smoother data than
expected given the errors. For a consistency check, we repeated the
fits with different numbers of parameters, namely $i=5,j=11$ and
$i=4,j=9$. In
figures~\ref{total_mass_density_different_parameters_sets} --
\ref{temperature_different_parameters_sets} we show the profiles of
the total mass density, the gas mass density, and the temperature for
the three free parameters sets, ($i=5,j=11$), ($i=6,j=14$) and
($i=4,j=9$). The figures clearly show that our results are insensitive
to the precise number of adopted free parameters. 
To assess the degree of correlation between values of the projected 
fit parameters we have calculated the correlation matrix for $i=4,j=9$, 
which is specified in table~\ref{correlation matrix} in the Appendix.

We tested the effect of the form of extrapolation to large radii on
the values obtained for the total mass density and the gas mass
density. We checked the results for large changes in the extrapolation
power-law index, i.e. $-2$ and $-4$ (compared to our standard
assumption of $\rho \propto r^{-3}$).  We used our standard number of
free parameters, $6$ and $14$ for the total mass density and for the
gas mass density, respectively. In each case the same extrapolation
index was used for the total mass and the gas mass density in order to
ensure a constant gas fraction at large radii. We expected a large
change in the free parameter at the largest radius, which is the
closest to the extrapolation. Using an extrapolation index $-2$ gave a
value which is lower by $21\%$ and $11\%$ for the total mass density
and gas mass density, respectively, at this last point. For an
extrapolation index $-4$ there was $<1\%$ change in the total mass
density value and $8\%$ in the gas mass density. In the gas mass
density there was also a change in the innermost radius, $29\%$ and
$36\%$ for an extrapolation index of $-2$ and $-4$, respectively; this
change is not very significant due to the large errors at this
radius. Taking a smaller number of free parameters lowers the errors
of the gas mass density at the smaller radius and as a consequence
also weakens the dependence on the extrapolation index. Thus, our
results in general are not strongly dependent on the extrapolation
index.

\subsection{Total mass density} 
\label{The total mass density}

The values of our six free parameters of the total mass density and
the deduced 3D mass are shown in table (\ref{total mass density
table}). As mentioned above the X-ray temperature data were not
used. The surface density data were used, but their effect on the
total mass profile is limited since $\rho_{\rm gas}$ appears on both
sides of equation~(\ref{eq: HEE}).  Thus, the lensing data basically
determined the derived 3D total mass density and the 3D mass profiles.

\begin{table}

\caption{Values of 3D mass density; errors are 1-$\sigma$ confidence. 
\label{total mass density table}}
\begin{center}
\begin{tabular}{|c|c|c|}

\hline
$r$ [h$^{-1}$ kpc] & $\rho_{\rm total}$ [$10^{-25}$ h$^{2}$
gr/cm$^3$] & $M_{\rm total}$ [$10^{12}$ h$^{-1}$ $M_{\sun}$]\\
\hline                    
 10.1               & $90_{-90}^{+91}$                              & $0.91   \pm 0.68$  \\
 27.7               & $39 \pm 11$                                 & $7.3    \pm 2.5$  \\
 75.9               & $10.5 \pm 1$                                    & $48.3   \pm 3.9$  \\ 
 208              & $2.4 \pm  0.4$                                  & $252  \pm 15$ \\  
 568              & $0.13\pm 0.07$                                  & $679  \pm 82$   \\
 1554              & $0.0022_{-0.0022}^{+0.0049}$                    & $956  \pm 180$\\
\hline
\end{tabular}
\end{center}
\end{table}

We also compared the total mass density profile obtained by the
model-independent method to the that obtained by fitting particular 
models. 
We tested the NFW profile,
\begin{equation}
\rho = \frac{\rho_0}{\left(r/r_s \right)\left(1+r/r_s \right)^2}\ ,
\end{equation} 
and a core model,
\begin{equation}
\rho = \frac{\rho_0}{\left( 1+(r/r_s)^2 \right)^n}\ ,
\end{equation} 
where $r_s$ is a scale radius. 
In the NFW profile $\rho_0$ and $r_s$ were free parameters, and in 
the core profile $\rho_0$, $r_s$, and $n$ were free parameters. 
\begin{table}
\caption{The values of the parameters of the two total mass density 
models, NFW and core. The errors are 1-$\sigma$ confidence. 
\label{profiles fit to the lensing data}}
\begin{center}
\begin{tabular}{|l|c|c|}
\hline
Parameter & NFW & core \\
\hline
$\rho_0$ [$10^{-25}$ h$^{2}$ gr/cm$^3$] &   $9.6\pm1.8$    & $24.4\pm4.5$   \\                
$r_s$ [h$^{-1}$ kpc]                             &   $175\pm18$ & $91\pm17$  \\
n                                       &   -                & $1.43\pm0.12$    \\  
$\chi^2/dof $                           &   $15.3/(26-2)$   & $ 13.3/(26-3)$  \\
\hline
\end{tabular}
\end{center}
\end{table}
We chose these two profiles since they are frequently used and are
significantly different both at small and at large radii. For simplicity,
we fitted the two profiles only to the lensing data. 
\begin{figure}
\centerline{\includegraphics[width=8.5cm]{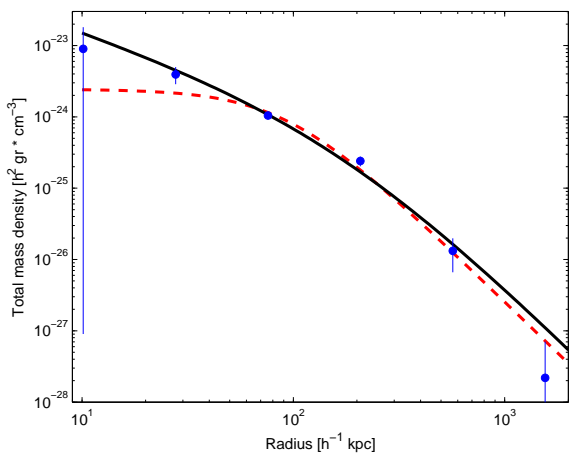}}
\caption{A comparison between the mass profile derived by the model-independent method 
to that using particular models. We compare the values obtained by our
model-free analysis (points with error bars) to the results of
assuming an NFW profile (solid curve) or a core plus power-law profile
(dashed curve). \label{free_param_Vs_NFW_and_core}}
\end{figure}

A comparison of the profiles to the six values we obtained in our main
analysis (figure~\ref{free_param_Vs_NFW_and_core} and
table~\ref{profiles fit to the lensing data}) yields a fairly close
agreement and implies that neither of the profiles is strongly
excluded. Still, the measured profile as sampled by the six points is
steeper at large radii than both models, in agreement with Broadhurst
et al.\ (2005b).  The core profile is too shallow at small radii and
thus seems not to be a good fit to our deduced values of the mass
density in this region.

The NFW fit gave a concentration parameter $C_N=12.2^{+0.9}_{-1}$,
where $C_N=r_{\rm vir}/r_s$ in terms of the virial radius $r_{\rm
vir}$ and the characteristic radius $r_s$ of the NFW profile. This is
close to the value obtained by Broadhurst et al.\ (2005b),
$C_N=13.7^{+1.4}_{-1.1}$, which was based only on a fit to the strong
and weak lensing information. We also obtained from the NFW fit two
characteristic radii for the cluster: $r_{200}$, the radius inside
which the average density is $200$ times the critical density, and
$r_{\rm vir}$, defined with a relative density of $116$, which is the
density expected theoretically at $z=0.183$ in the $\Lambda$CDM model.
The fit yielded $r_{200}=1.71$ h$^{-1}$ Mpc and $r_{vir}=2.14$
h$^{-1}$ Mpc.

\begin{figure}
\centering
\epsfig{file=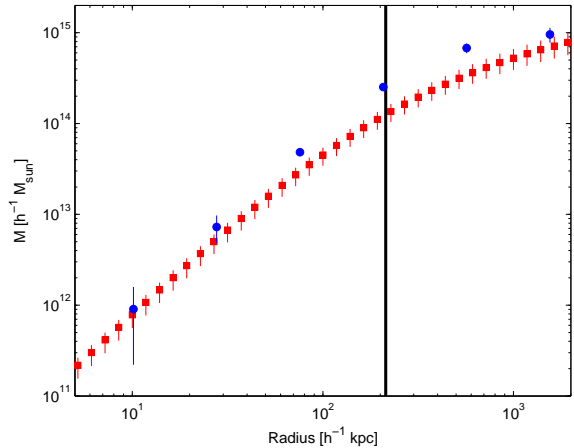, width=8.5cm, clip=}
\caption{The derived 3D total mass profile. We compare our derived
profile (dots) to the one derived by AM04 (squares). They used an NFW
profile based on data out to $693$ h$^{-1}$ kpc. Both profiles are
shown with 1-$\sigma$ errors. The vertical dotted line is at 0.1
$r_{\rm vir}$.\label{total mass profile}}
\end{figure}

In figure~\ref{total mass profile} we plot our derived 3D total mass
profile from eq.~(\ref{eq: M}) and the one derived in AM04 from XMM
data. They used an NFW profile based on data out to $693$ h$^{-1}$
kpc. The two profiles agree at the small radii but disagree at larger
radii. Our highest-radius point agrees with their profile, but at this
radius we had to extrapolate the AM04 3D mass profile beyond their
data range.

\subsection{2D mass profile}

\begin{figure}
\centering
\epsfig{file=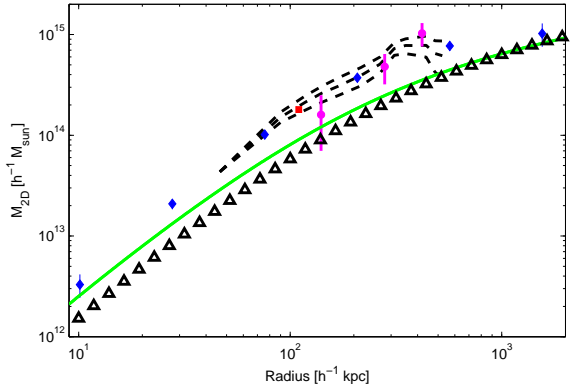, width=8.5cm, clip=}
\caption{The derived 2D mass profile (6 diamonds) is compared to the
gravitational lensing results from strong lensing (square, Tyson \&
Fischer 1995), distortion of background galaxy luminosity function (3
circles, Dye et al.\ 2001), deficit in number counts of red background
galaxies (dashed curves, Taylor et al. 1998), projected best-fit NFW
model from weak gravitational shear (triangles, King et al.\ 2002),
and projected best-fit NFW model from X-ray data (solid curve, AM04) .
\label{M2D comparison}}
\end{figure}

The deduced 2D mass profile $M_{2D}(R)=2\pi \Sigma_{crit} \int_{0}^{R}
\kappa(R^\prime) R^\prime dR^\prime$, is compared with previous
results in Figure~\ref{M2D comparison}.  These include results from
strong lensing (Tyson \& Fischer 1995), the lensing magnification
measured by the distortion of the background galaxy luminosity
function (Dye et al.\ 2001), the lensing magnification measured by the
deficit of red background galaxies (Taylor et al.\ 1998), the
projected best-fit NFW model from X-ray data (AM04 via XMM), and the
best-fit NFW model from weak gravitational shear analysis (Clowe \&
Schneider 2001; King et al.\ 2002). The latter analysis, although it
is a weak lensing analysis, is the closest to the profile derived from
X-ray data, but it appears to have underestimated the distortion
signal derived in the other lensing analyses.  This is probably caused
by confusion between the cluster galaxies and the foreground or
background galaxies. Thus, the X-ray analysis of AM04 (based on the
X-ray temperature) gives a substantially lower 2D mass than the more
reliable lensing analyses. This foreshadows the discrepancy that we
find with the measured temperature (see section~5.5 below).

\subsection{Gas density profile}

The values of the derived gas mass density are shown in table~\ref{gas
mass density table}. We compared our gas density profile to the one
obtained by MME99. They fitted a single $\beta$ model to the surface
brightness profile; in clusters where they suspected the presence of
cooling flows the fit was to a double $\beta$ model (though with the
same $\beta$ in both elements).  They used two types of clues for the
presence of cooling flows. First, when fitting a single $\beta$ model
the surface brightness data points must be higher than the fitted
curve in the inner region. Second, the cluster must appear relaxed,
i.e., lacking obvious asphericity or substructure that would indicate
a recent merger.  Judging by their successful fit to a single $\beta$
model, they concluded that A1689 is not likely to have a cooling flow.
Their fit is thus based on an assumed isothermal profile, and $0.3$
solar abundance. Figure~\ref{gas density comparison to MME99} shows
excellent agreement between our obtained gas density profile and the
one obtained by MME99.
\begin{table}

\caption{Deduced values of the 3D gas mass density, and derived
temperatures using the gas and total mass profiles together with the
hydrostatic equilibrium equation.  The errors are 1-$\sigma$
confidence. \label{gas mass density table}}
\begin{center}
\begin{tabular}{|c|c|c|}

\hline
$r$ [h$^{-1}$ kpc]   & $\rho_{\rm gas}$ [$10^{-26}$ h$^{0.5}$ gr/cm$^3$] & $ T_{3D}$ [keV] \\
\hline                    
12.9                 & $9.9_{-9.9}^{+30.3}$                       & $10.1_{-10.1}^{+30.5}$        \\
17.6                 & $7.2_{-7.2}^{+10.7}$                       & $12.7_{-12.7}^{+15.5}$         \\
24.0                   & $6.4\pm3.1$                                 & $13.0\pm4.7$                   \\ 
32.7                 & $5.06\pm0.55$                                & $14.0\pm1.8 $                  \\  
44.5                 & $4.12\pm 0.18$                                & $14.1\pm1.0 $                  \\
60.7                 & $3.09\pm0.11$                                & $14.7\pm0.75$                   \\
82.7                 & $1.96\pm0.055$                               & $17.1\pm0.7$                   \\
113                & $1.35\pm0.033$                               & $17.8\pm0.7$                    \\
154                & $0.97\pm0.023$                               & $16.65\pm0.78$                   \\
209                & $0.567\pm0.015$                              & $16.94\pm0.96 $                  \\
285                & $0.35\pm 0.012$                               & $14.8\pm1.0$                   \\
388                & $0.19\pm0.0079$                              & $13.1\pm 1.3$                  \\
529                & $0.0982\pm0.00067$                           & $11.3\pm1.7$                   \\
721                & $0.0586\pm0.00057$                            & $7.29\pm0.87$                    \\

\hline
\end{tabular}
\end{center}
\end{table}

\begin{figure}

\centering
\epsfig{file=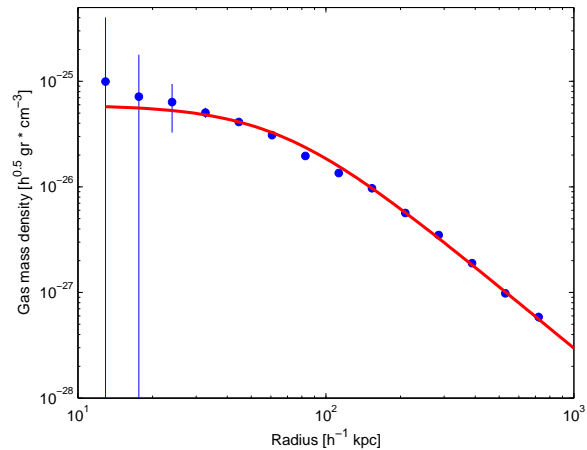, width=8.5cm, clip=}
\caption{Our derived gas mass density profile (points with error bars) compared to the
profile derived by MME99 (solid curve).\label{gas density comparison to
MME99}}
\end{figure} 

\begin{table}

\caption{Deduced values of the 3D gas mass at various radii. 
The errors are 1-$\sigma$ confidence. \label{gas mass table}}
\begin{center}
\begin{tabular}{|c|c|}

\hline
$r$ [h$^{-1}$ kpc]   & $M_{\rm gas}$ [$10^{11}$ h$^{-5/2}$ $M_{\sun}$]\\
\hline                    
12.9                 & $0.20_{-0.20}^{+0.60}$ \\
17.6                 & $0.37_{-0.37}^{+0.70}$  \\
24                   & $0.71 \pm 0.57$         \\ 
32.7                 & $1.44  \pm 0.68$         \\  
44.5                 & $2.93  \pm 0.64$          \\
60.7                 & $5.85  \pm 0.66$          \\
82.7                 & $10.93 \pm 0.65$          \\
113                & $19.46 \pm 0.66$          \\
154                & $34.63 \pm 0.67$          \\
209                & $59.3  \pm 0.74$          \\
285                & $96.83 \pm 0.88$          \\
388                & $151.5\pm 1.3$          \\
529                & $224.8\pm2.5$           \\
721                & $328.4\pm3.7$           \\
\hline
\end{tabular}
\end{center}
\end{table}

\begin{figure}
\centering
\epsfig{file=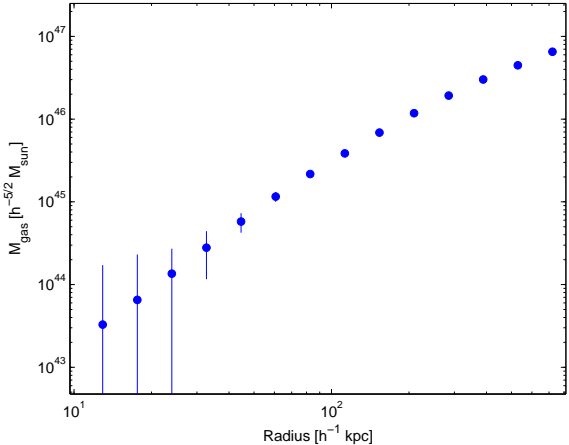, width=8.5cm, clip=}
\caption{The gas mass profile derived from the best-fit gas density
profile. All the values have error bars but some are too small to
distinguish. These values are also listed in table~(\ref{gas mass
table}).}
\end{figure}

\subsection{Gas mass fraction}

Having derived the total and gas mass densities, the gas fraction,
$f_{\rm gas}=M_{\rm gas}/M_{\rm total}$, can then be determined upon
integration.  We can compare our value, $f_{\rm
gas}(r=0.25r_{200})=0.0326\pm0.0023$ h$^{-3/2}$, to Sanderson et al.\
(2003). They obtained for the $8-17$ keV range (see their figure~6)
$f_{\rm gas}(r\cong0.25r_{200})= 0.041$ h$^{-3/2}$. Our value of
$r_{200}$ is slightly different since it was derived from the NFW fit
to the lensing data, whereas their value was deduced from the X-ray
data. They obtained $r_{200}=2955$ h$_{0.7}^{-1}=2068.5$ h$^{-1}$ kpc,
and if we scale our result by their $r_{200}$ we get $f_{\rm
gas}(r=0.25r_{200})=0.0359\pm0.0024$ h$^{-3/2}$. This value is still
somewhat lower than their $8-17$ keV value, but in good agreement if
we look at the scatter of the gas fraction in their figure~4. In their
figure~4 they plotted the values of the gas fraction of different
clusters at $r=0.3r_{200}$ as a function of the cluster
temperature. The gas fraction of hot clusters ranges from $\sim 0.035$
h$^{-3/2}$ to $\sim 0.06$ h$^{-3/2}$, with some $\sim40 \%$
uncertainty.  AM04 found for A1689 $f_{\rm
gas}(r=r_{2500})=0.044\pm0.0049$ h$^{-3/2}$, which is in agreement
with our value of $f_{\rm gas}(r=r_{2500})= 0.0385\pm0.0026$
h$^{-3/2}$. MME99 found for an ensemble of clusters with $T>5$ keV a
$f_{\rm gas}(r=r_{500})=0.075$ h$^{-3/2}$, close to and slightly lower
than the value in Sanderson et al.\ (2003), just as our gas fraction
at $r=0.25r_{200}$ was a little lower than the value derived by
Sanderson et al.\ (2003) at that radius.

Markevitch et al.\ (1999) examined two relaxed clusters, A2199 and
A496, which are cooler than A1689 and have temperatures $\sim 4-5$
keV. They found $f_{\rm gas}(r=0.5\;h^{-1}\;{\rm
Mpc})=0.0569\pm0.0049$ h$^{-3/2}$ and $f_{\rm gas}(r=0.5\;h^{-1}\;{\rm
Mpc})=0.0559\pm0.006$ h$^{-3/2}$ for A2199 and A496, respectively.
These values are slightly higher than in other papers. They claimed
that their values are consistent with others including MME99, who
found that $\langle f_{\rm gas} \rangle =0.0566$ h$^{-3/2}$ for
clusters with $T<5$ keV.  However, note that the Markevitch et al.\
values are at $r_{1000}$ while the MME99 values are at $r_{500}$.

The gas mass fraction was also determined from SZ measurements with
BIMA and OVRO (Grego et al.\ 2001). The latter authors measured
$f_{\rm gas}(r=65")$ and used a scaling relation in order to obtain
$f_{\rm gas}(r=r_{500})$. They found for a sample of 18 clusters
(assuming $\Omega_{m}=0.3$, $\Omega_{\Lambda}=0.7$) $f_{\rm
gas}=0.081^{+0.009}_{-0.011}$ h$^{-1}$. These values are higher than
obtained by X-ray measurements. For A1689, Grego et al.\ (2001)
obtained $f_{\rm gas}(r=r_{500})=0.098^{+0.029}_{-0.033}$ h$^{-1}$, or
more directly $f_{\rm gas}(r=65")=0.068^{+0.02}_{-0.023}$
$h^{-1}$. Our results yield $f_{\rm gas}(r=r_{500})=0.0482$ h$^{-3/2}$
and $f_{\rm gas}(r=65"=140\;h^{-1}\;{\rm kpc})=0.0222\pm0.0009$
h$^{-3/2}$, much lower than in their paper. Some systematic
discrepancy in the gas fraction from X-ray and SZ measurements is
expected (Hallman et al.\ 2007). The difference can be due to the fact
that $M_{\rm gas}({\rm SZ})\propto T^{-1}$ (where $T$ is the gas
temperature as deduced from spectral X-ray measurements; Grego et al.\
2001), since (the thermal component of) the SZ effect depends on the
product of gas density and temperature.  Thus, an underestimation of
the temperature (see section~\ref{The 2D temperature profile section})
results in an $M_{\rm gas}({\rm SZ})$ that is higher than $M_{\rm
gas}(\mbox{X})$, making $f_{\rm gas}({\rm SZ})$ higher than $f_{\rm
gas}(\mbox{X})$.  Expressing the difference in terms of the gas mass,
Grego et al.\ (2001) obtained $M_{\rm
gas}(r=65")=4.6_{-1.1}^{+0.8}\cdot 10^{12}$ h$^{-2}$ $M_{\odot}$ and
we $M_{\rm gas}(r=65")=2.9\cdot 10^{12}$ h$^{-1}$ $M_{\odot}$. The
ratio between the gas fraction derived by X-rays and the one derived
by SZ also depends on clumpiness, where we define the clumping
parameter as $C\equiv \langle n_{e}^{2} \rangle ^{1/2}/ \langle n_{e}
\rangle$. The gas mass measured via X-ray is $M_{\rm
gas}(\mbox{X})\propto C^{-1}$, i.e., the actual gas mass is lower than
that apparently observed if the cluster is clumpy but this is not
corrected for. The gas mass measured via SZ is not proportional to the
clumping. Thus, correcting for clumping would further decrease $M_{\rm
gas}(\mbox{X})/M_{\rm gas}({\rm SZ})$ by a factor of $\sim C$. This
may indicate that the level of clumping in this cluster is low.

\begin{figure} 
\centering
\epsfig{file=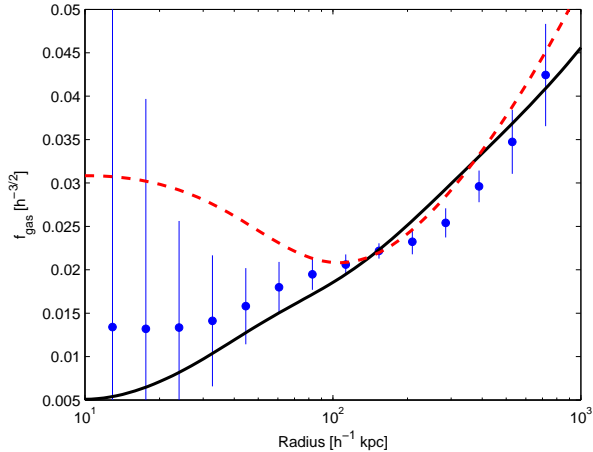, width=8.5cm, clip=}
\caption{A comparison between the $f_{\rm gas}$ profile derived by our
model-independent method (points with error bars) to the profile
derived using parametrized models. We consider two simple models, NFW
(solid curve) and a core profile (dashed curve) for the total mass, in
each case with a double beta profile for the gas mass.
\label{f_gas comparison}}
\end{figure}

Finally, we also compared our results for the gas fraction to the
simple theoretical models. In figure~\ref{f_gas comparison} we show
the $f_{\rm gas}$ obtained by the parametrized models, NFW (solid
curve) or core (dashed curve) for the total mass density and (in both
cases) a double $\beta$ model for the gas mass density, compared to
the profile obtained from our model-free method (points with error
bars). In general the models are not far from our reconstructed
values. At very large radii, extrapolated beyond the data points, the
gas fraction derived by the models would diverge from the real,
supposedly constant value at large radii, since $\rho_{\rm gas}\propto
r^{~-2.2}$ and $\rho_{DM/total}\propto r^{-3}$. Instead, in our
model-independent method the extrapolation of the gas density is
$\rho_{\rm gas}\propto r^{-3}$, in order to give an asymptotic $f_{\rm
gas}=$const at large radii. At radii of $\sim200 - 500$ h$^{-1}$ kpc
the $f_{\rm gas}$ obtained by the model-independent method tends to be
slightly lower than the one obtained by the parametrized models. At
small radii the NFW model fits better than the core model, though the
core gives a better overall fit to the lensing data (see
table~\ref{profiles fit to the lensing data}).

\subsection{2D temperature profile} 
\label{The 2D temperature profile section}

\begin{figure}
\centering
\epsfig{file=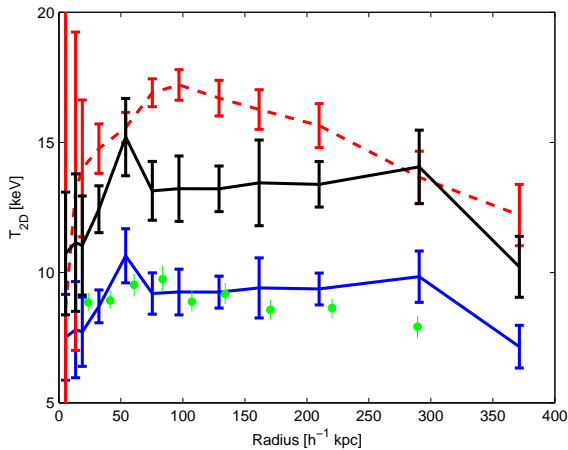, width=8.5cm, clip=}
\caption{A comparison between the 2D temperature profile obtained 
from our model-independent method (dashed curve) to the profile
measured from the X-ray spectrum (lower solid curve). Also shown are
the measured temperature divided by 0.7 (upper solid curve) and the
measured 2D temperature profile from AM04 (circles with error
bars). \label{T2D comparison}}
\end{figure}

\begin{figure}
\centering
\epsfig{file=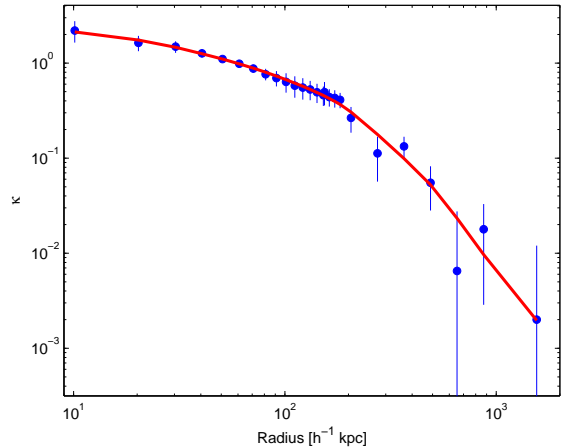, width=8.5cm, clip=}
\caption{Our fit to the lensing data. The $\kappa$ profile obtained by
our model-independent method (solid curve) is compared to the measured
profile (points with error bars), where $\kappa$ is the 2D surface
density in units of the critical density for lensing.
\label{lensing fit}}
\end{figure}

\begin{figure}
\centering
\epsfig{file=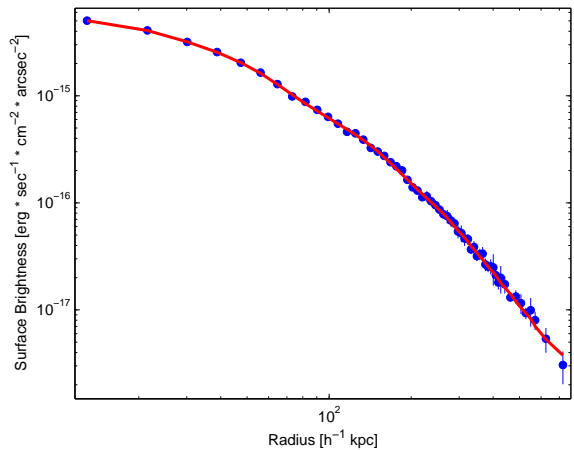, width=8.5cm, clip=}
\caption{Our fit to the surface brightness data. The X-ray surface
brightness profile obtained by the model-independent method (solid
curve) is compared to the measured profile (points with error bars).
\label{surface brightness fit}}
\end{figure}

In figure~\ref{T2D comparison} we compare the 2D temperature obtained
by our model-free method to the measured 2D temperature. The measured
2D temperature is substantially lower. It has been found in numerical
simulations that the spectroscopic temperature is typically lower than
the one determined by hydrostatic equilibrium by a factor of $0.7-0.8$
(Rasia et al.\ 2005). Thus we also show in the figure the measured 2D
temperature divided by 0.7 (upper solid curve). This factor explains
most of the discrepancy in A1689. Kawahara et al.\ (2006) used
cosmological hydrodynamical simulations to explain this
discrepancy. They found that local inhomogeneities in the gas are
largely responsible.

\begin{figure}
\centering
\epsfig{file=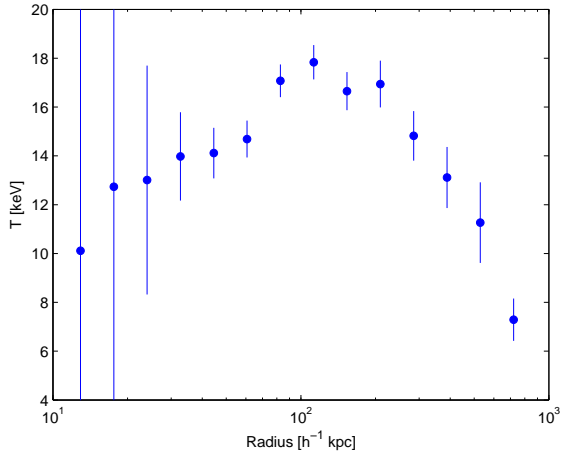, width=8.5cm, clip=}
\caption{The 3D temperature profile as reconstructed by our model-independent 
method. \label{T3D profile}}
\end{figure}

\subsection{The entropy and polytropic index}

Previous theoretical work has shown that the entropy profiles of
non-radiative clusters approximately follow a power law with $K(r)
\propto r^{1.1}$ (Tozzi \& Norman 2001; Borgani et al.\ 2002; Voit et
al.\ 2003).  More recent detailed hydrodynamical simulations by Voit,
Kay \& Bryan (2005) confirmed the power law behavior of the entropy at
$r>0.1r_{\rm vir}$. They also determined the normalization $N$, where
$K(r)/K_{200}=N(r/r_{200})^{1.1}$, which was found to be $1.32\pm0.03$
and $1.41\pm0.03$ for SPH and AMR simulations, respectively.  In the
central region of the cluster, $<0.1r_{\rm vir}$, there are some
claims for the existence of an entropy "floor", i.e., a limiting
constant value.  Voit et al.\ (2003) showed theoretically how an
entropy floor can be attained; Voit, Kay \& Bryan (2005) used
simulations to find the level of this entropy floor. They found an
entropy floor using an AMR code, but the value obtained with an SPH
code was substantially lower (if at all present).

Observational comparisons of the entropy of clusters at small and at
large radii show a departure from the predicted self-similarity. In
the inner regions of low-mass clusters there appears to be a "floor"
in entropy of $\sim 135$ keV cm$^2$, with a higher value for high-mass
clusters (Lloyd-Davies, Ponman, \& Cannon 2000). This may represent a
"preheated" minimum level which Kaiser (1991) speculated may be due to
the effect of star formation in early galaxies which preheat the IGM
through galactic winds. This idea is strengthened by the ubiquitous
presence of gas outflows in observations of high-redshift galaxies
(Franx et al.\ 1997, Frye \& Broadhurst 1998, Frye, Broadhurst \&
Benitez 2002, Pettini et al.\ 2001). Thus it seems natural to link
this entropy floor with the winds from galaxy formation.

\begin{figure}
\centering
\epsfig{file=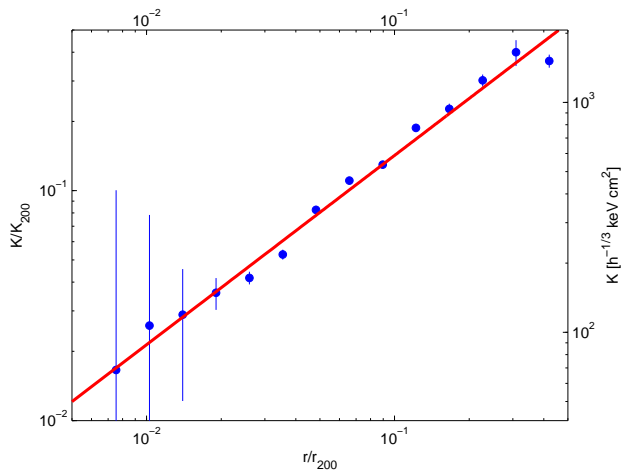, width=8.5cm, clip=}
\caption{Our derived entropy profile (points with error bars) compared
to the best-fit power law profile (solid curve).
\label{entropy profile}}
\end{figure}

In figure~\ref{entropy profile} we plot $K/K_{200}$, where $K_{200}$
is the entropy at $r=r_{200}$. We find $r_{200}= 1.7$ h$^{-1}$ Mpc and
$r_{200}\cong0.8 r_{\rm vir}$.  The value of our derived entropy at
$0.1r_{\rm vir}$ agrees well with the one found by Lloyd-Davis,
Ponman, \& Cannon (2000), who used ROSAT and ASCA data.  They assumed
spherical symmetry and hydrostatic equilibrium, and a single beta
model for the gas density along with a linear function for the
temperature. For the DM density they used an NFW profile. Our derived
entropy at $0.1r_{\rm vir}$, $K(r=0.1r_{\rm vir})=786\pm33$ h$^{-1/3}$
keV cm$^2$, also agrees well with Ponman, Sanderson, \& Finoguenov
(2003), who also used ROSAT and ASCA data, assumed spherical symmetry
and hydrostatic equilibrium, a single beta model for the gas density,
and a linear function or a polytrope for the temperature. We fitted a
power law to our entropy profile at all radii where we obtained the
gas mass density (which includes points at $r<0.1r_{\rm vir}$).  This
fitted power law, which is shown in figure~\ref{entropy profile}, is
$K(r)/K_{200} = (0.95\pm 0.05)(r/r_{200})^{0.82\pm 0.02}$.

Our X-ray data go up to $\sim0.34r_{\rm vir}$, so obtaining a fit to
the $(0.1-1)r_{\rm vir}$ range is not possible. In the available range
$\sim (0.006-0.34)r_{\rm vir}$ the power law is a close fit, though
the index is somewhat lower than the theoretical value of $1.1$. 
Excluding the entropy points at small radii, $r<0.1r_{\rm vir}
$, would not substantially change the value of the power index, but
then there are only four data points at $r>0.1r_{\rm vir}$. In the
inner part ($r<0.1r_{vir}$) if the entropy is flattened then this
occurs only at $r<0.02r_{vir}$. This is consistent with Tozzi \&
Norman (2001) who found that the flat entropy region in massive
clusters ($M \sim 10^{15}$ $M_{\odot}$) occurs only at $r<0.01\;
r_{200}$. Comparing our obtained values of the entropy at $r<0.1r_{\rm
vir}$ to those in Voit, Kay \& Bryan (2005), they are closer to the
values obtained in the SPH simulations than in the AMR simulations. 
Our derived normalization $0.95\pm 0.05$ is somewhat lower than obtained 
by Voit, Kay, \& Bryan (2005) in both types of simulations.

The deduced power law index is slightly lower than the value obtained
from simulations (Tozzi \& Norman 2001; Borgani et al.\ 2002; Voit et
al.\ 2003; Voit, Kay, \& Bryan 2005). The disagreement can be
explained by the fact that we did not assume an NFW profile. Assuming
an NFW profile gives $K(r)/K_{200} = (1.13\pm 0.39 )(r/r_{200})^{1\pm
0.2 }$.  This is in good agreement with the "expected" power law index
of $1.1$. There is also a good agreement with the normalization
obtained by simulations (Voit, Kay, \& Bryan 2005). It has also been
shown in simulations that as the amount of preheating increases, the
power law index becomes flatter at radii beyond the flat inner region
(Borgani et al.\ 2005).

\begin{figure}
\centering
\epsfig{file=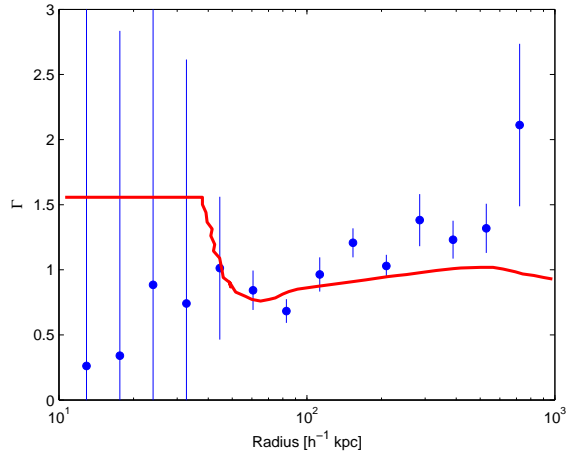, width=8.5cm, clip=}
\caption{The adiabatic index deduced in this work (points with error bars),  
and the profile derived by Tozzi \& Norman (2001) (solid curve). 
\label{politropic index}}
\end{figure}

In figure~\ref{politropic index} we plot the adiabatic index. The
solid curve is taken from Tozzi \& Norman (2001). They used an NFW
profile for the DM and assumed no cooling. They also assumed an external, 
initial adiabat for the entropy. Our derived adiabatic index profile is 
in reasonable agreement with the profile derived by Tozzi \& Norman (2001).

\subsection{The effect of the cD galaxy}

Figures~\ref{T2D comparison} and \ref{T3D profile} show that the 2D
and 3D temperature profiles peak at $r\sim 0.1r_{500}$ and decline at
large radii. At smaller radii the temperature profile also declines.
This decline of the temperature profile at small radii occurs within
the region that includes emission from the cD galaxy.  Since the gas
in the cD is denser and colder than the cluster gas, it might be the
cause for this decline at small radii. In order to evaluate the effect
of the cD on the cluster we modeled the emission of the cD. We
estimated the gas mass density profile by taking the profile of M87,
which is the cD in the nearest cluster of galaxies. We thus fitted a
double $\beta$ model (as Matsushita et al.\ 2002 did) to the surface
brightness data in M87. The values of the best-fitted parameters of
the gas number density of the cD are shown in table~\ref{parameters of
the cD gas number density double beta model}. Our obtained values
agree well with the ones in Matsushita et al.\ (2002). For the
evaluation of the cD mass profile we took $R_e=7.8$ h$^{-1}$ kpc,
$r_s=10$ h$^{-1}$ kpc, $M_{\rm cD}=10^{13}$ $M_{\odot}$, and $M_{\rm
stars}=5\cdot 10^{11}$ $M_{\odot}$ (for the mass profile see
section~\ref{sec: cD}).

\begin{table}
\caption{The values of the best fitted parameters of the gas number density of 
the cD assuming a double beta model for the gas mass density of cD. 
We fitted to the values taken from Matsushita et al.\ (2002)
\label{parameters of the cD gas number density double beta model}}
\begin{center}
\begin{tabular}{|l|c|}
\hline
Parameter                  & Value                             \\
\hline
$n_e(1,0)$   [$cm^{-3}$]   & $0.117^{+0.016}_{-0.014}$         \\
$r_{c,1}$    [arcmin]      & $0.366^{+0.071}_{-0.064}$         \\
$\beta_{1}$                & $0.439^{+0.033}_{-0.025}$         \\
$n_{e,2}(0)$ [$cm^{-3}$]   & $6.3^{+0.8}_{-0.7}\cdot 10^{-3}$  \\
$r_{c,2}$    [arcmin]      & $5.41^{+0.31}_{-0.33}$            \\
$\beta_{2}$                & 0.47(fixed)                       \\          
\hline
\end{tabular}
\end{center}
\end{table}

\begin{figure}
\centering
\epsfig{file=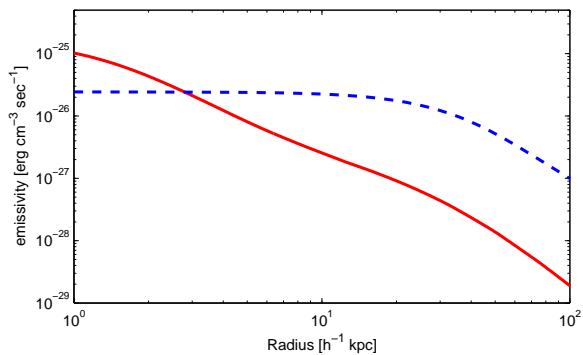, width=8.5cm, clip=}
\caption{A comparison between the emissivity profiles of the cluster
(dashed line) and of the cD galaxy (solid line).  The cD emissivity
profile is derived by assuming it is similar to that of M87.
\label{emissivity cD Vs cluster}}
\end{figure}

We checked the innermost part of the cluster $r\lesssim 10$ h$^{-1}$
kpc. This is out of the data range and requires an extrapolation, so
we extrapolated using particular profiles. We used NFW for the DM
profile and a double $\beta$ model for the gas profile. In
figure~\ref{emissivity cD Vs cluster} we plot the emissivity of the cD
galaxy and compare it to the emissivity of the cluster. The figure
shows that the cD is dominant only at $r<4$ h$^{-1}$ kpc. Since the
innermost surface brightness data point is at $4$ h$^{-1}$ kpc, the
effect of the cD galaxy is likely minor, as expected in a rich cluster
like A1689.

\begin{figure}
\centerline{\includegraphics[width=8.5cm]{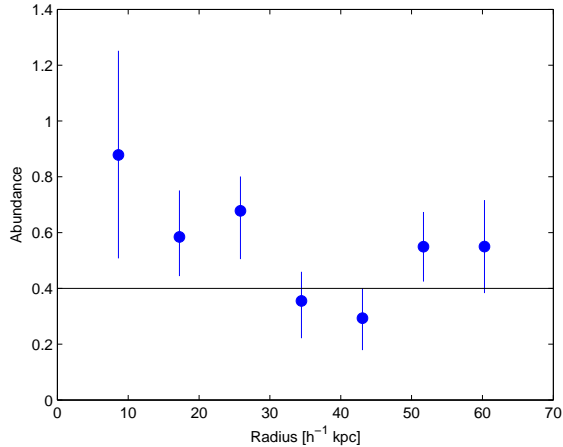}}
\caption{A comparison of the abundance (in solar units) within 4"
annuli (points with error bars) to the average abundance of the
cluster in a $3^\prime$ aperture (solid line). The errors may be
underestimated (see text).
\label{inner profile of the abundance} }
\end{figure}

We checked for an indication in the abundance profile that the cD
contributes only at $<4$ h$^{-1}$ kpc (if at all). Note that the
spatial resolution of Chandra is $\sim 0.5"$, which for the redshift
of A1689 is $\sim 1.5$ h$_{0.7}^{-1}$ kpc. Also the cD galaxy
coincides with the X-ray centroid within $\sim 1.5"$ uncertainties
(XW02). Figure~\ref{inner profile of the abundance} indicates that the
abundance may be higher, $\sim 0.9$, in the core than in the outer
region.  However, this is not statistically significant due to the
narrow annuli used which give high errors. The errors were calculated
here after freezing the temperature and the normalization. If these
parameters are allowed to vary, this will of course increase the
errors in the abundance.
As expected from the low emissivity of the cD galaxy compared to that
of the cluster, it is hard to see spectroscopically an indication of
the cD. We fitted a double temperature model, WABS(MEKAL+MEKAL), to
the $3^\prime$ aperture, where the abundance of the first component
(the cD galaxy) was fixed to solar and the second (the cluster) to 0.4
solar (the average value in the cluster). We also fixed the Galactic
absorption to be $2\cdot 10^{-20}$ cm$^{-2}$ as appropriate for this
direction (Dickey \& Lockman 1990).  The resulting best fit is with
two temperatures, a cold one $\sim 0.5$ keV and a hot one $\sim 10$
keV. This, however, might be due to the real multi-temperature nature
of the cluster gas. Extracting a smaller region, where the cD flux
should be less diluted by the cluster flux, gave about the same two
temperatures but with high errors. Thus, spectroscopic fitting was
inconclusive.

\section{Discussion}

The increasing quality of X-ray and optical imaging
data motivates renewed and more thorough examinations of the physical
nature of galaxy clusters as revealed by the very different processes
of bremsstrahlung radiation and the gravitational deflection of
light. In this paper we have examined the apparently relaxed cluster 
A1689, where only minimal substructure is evident from the dark matter,
galaxy, and X-ray distributions and where the X-ray emission is smooth 
and symmetric. The longstanding claims of discrepancies in the total 
cluster mass estimated from these different kinds of analysis can now be
investigated with greater precision, fewer assumptions, and in a more
model-independent way.

There is a strong incentive to describe in analytical form the general
nature of the total mass, gas, and temperature profiles. These
quantities are inextricably bound up with the process of structure
formation, including the nature of dark matter and the cooling history
of the gas including interaction and merging of substructure. Thus,
having an analytical form of the profiles is very useful.  The most
commonly examined profile for the dark matter, based on N-body
simulations of collisionless dark matter, is the NFW profile. In
contrast the most commonly fitted gas mass density or surface
brightness profile, the $\beta$ model, is essentially empirically
based. The temperature profile is often derived from the analytical
expression of the gas density using the polytropic equation of state,
or the gas is taken to be isothermal since the temperature gradient
was undetectable with the old X-ray satellites. This way of analyzing
the cluster can lead to substantial errors in the DM density, gas
density, and temperature values and the derived quantities such as the
overall mass.

As we discussed in the Introduction, the total mass profile can 
be independently determined from lensing measurements, or from X-ray 
measurements of the gas density and temperature profiles based on 
the assumption of hydrostatic equilibrium. (Clearly, the latter can 
be done also with spatially resolved S-Z measurements, when available.) 
The second basic assumption adopted in our analysis is spherical symmetry. 
Obviously, elongation along the line of sight is possible, but - as 
we have mentioned in the Introduction - this typically can introduce 
a 20\% bias in the mass estimate. For a detailed discussion of the 
impact of triaxial cluster morphology, see Gavazzi (2005). 

We have suggested here a model-independent method which uses free 
parameters and does not assume a specific profile for any mass 
component. The only hidden assumption is a linear (in log-log) 
interpolation between the free parameters; this just assumes a 
reasonable degree of smoothness in the profiles. The profiles 
are also extrapolated beyond the data, but we showed that the 
results are insensitive to the detailed extrapolation. We 
specifically used a simple power law extrapolation
for both the total mass density and for the gas mass density, with the
same power law index assumed in order to approach a constant gas
fraction at large radii. Our model-independent method is the best way
to obtain the values of the important parameters of clusters at radii
where there are good quality data. Our model-independent method can and 
will be applied to joint analyses of measurements of other clusters.

Within the data range, the highest value of the 3D temperature
(fig.~\ref{T3D profile}) is 2--3 times higher than the lowest
value. Specifically, the temperature profile peaks at $r\sim
0.1r_{500}$ and declines at both smaller and larger radii. The denser
environment at the center of a cluster should naturally cause a
decrease in the temperature there. We indeed find in this cluster that
the 2D temperature decreases towards the center of the cluster, at $r<
50-100$ $h^{-1}$ kpc (figure~\ref{T2D comparison}). This can be due to
several mechanisms. First, the high gas number density in the center
of the cluster causes a rapid loss of energy and a decrease in the
temperature. As pressure support decreases, the gas should gravitate
towards the bottom of the cluster potential well, i.e., the center of
the cluster, in a so-called "cooling flow" (Fabian 1994).  A cooling
flow should manifest in a detectable surface brightness enhancement in
the X-ray emission from the central region of a cluster, $\sim
100-200$ kpc. Many studies have failed to find these cooling flows
(MME99), at least not as expected in a simple or a standard cooling
flow model (Fabian et al.\ 2001; Peterson et al.\ 2001). An
explanation for this failure can be a more complicated mechanism which
maybe includes reheating of the gas (e.g., by energetic particles;
Rephaeli \& Silk 1995), mixing, differential absorption, efficient
conversion of cooling flow gas into low-mass stars, an inhomogeneous
metallicity distribution (which is not the situation in A1689, at
least not in terms of a radial gradient: fig.~\ref{Abundance
gradient}), or disruption of cooling flows by a recent subcluster
merger (see discussion in Fabian et al.\ 2001; Peterson et al.\ 2001).

There are uncertainties about the criteria for finding cooling flows,
and even with fixed criteria the decision whether cooling flows exist
is still data dependent.  At times the "solution" has been to exclude
the inner, problematic region.  Finding the value of the gas mass
density in the model-independent approach bypasses these two problems,
the unknown physical processes and the data dependence, as it allows
for an analysis with fewer prior assumptions. Another way to reduce
the core temperature is through the effect of the cD galaxy that is
often anchored at the cluster center.  The cD galaxy has a lower
temperature and is denser than the cluster so it can in principle be
the main cause for the low temperature at the center.  However, based
on our our initial, limited study it seems unlikely that the cD has a
major effect in A1689, since the emissivity of the cD dominates that
of the cluster only below $r=4$ h$^{-1}$ kpc (fig.~\ref{emissivity cD
Vs cluster}).

We have found a good fit to the observed 2D profiles of the lensing
surface density and the X-ray surface brightness. Due to the
smoothness of the profiles, we were able to fit the data with a
relatively small number of free parameters. We have found a good
agreement with previous results for all the parameters we have checked
for A1689, including gas mass density, gas fraction profiles, measured
temperature, adiabatic index, and abundance (except compared to
AM04). The total mass density profile we obtained was essentially
determined directly by the lensing data alone, with $\lesssim 1\%$
differences introduced by using the X-ray data as well. This is the
case since we did not use the X-ray temperature as part of the fit.

We have shown that it is possible to obtain a model independent 3D
mass profile for which very good agreement is found between the
lensing mass profile and the X-ray emission profile. However, there
is still a discrepancy between the temperature derived directly from
(X-ray) measurements, and that deduced from a solution to the HEE,
with the latter $\sim30\%$ higher at all radii, as has already been
determined by Mazzotta et al.(2004) and Rasia et al. (2005). Other
phenomena, such as bulk motions, turbulence, and nonthermal degree 
of freedom may contribute to the pressure, and thus reduce the 
temperature with respect to the one obtained from assuming that 
thermal gas pressure is the only contributor. Specifically, 
Faltenbacher et al. (2005) claim that based on their high-resolution 
cosmological simulations - in which they identified and analyzed eight 
clusters at $z=0$ - about 10\% of the total pressure support may be 
contributed by random gas bulk motions, which may affect temperature 
by up to 20\%. It has also been suggested that X-ray luminous clumps 
of relatively low temperature may bias projected temperature 
measurements downward (Kawahara et al.\ 2006). In any case, as we have 
already mentioned this temperature discrepancy is much smaller than 
the mass discrepancy by a factor of $2-4$ encountered in previous 
(separate X-ray and lensing) analyses.

The derived entropy profile for A1689 has a power law form with no
obvious central flattening, at least at $r>0.02r_{\rm vir}$, as
expected for massive clusters from theoretical work (Tozzi \& Norman
2001). The existence and the level of the entropy floor is still not
completely clear in simulations, since different simulation methods
give different results (Voit, Kay, \& Bryan 2005). This is probably
due to the limited resolution of simulations which is especially
critical in the core of the cluster. Upcoming simulations with spatial
resolution of $\sim 5$ kpc should give a better understanding of the
entropy floor. Our derived radial slope is around $0.8$ rather than
$1.1$, flatter than the prediction. Since using simple parametrized
models for the DM and the gas gave a power law index of $1.1$, we
believe the value of $0.8$ might be a more accurate result, which
might be a reflection of a more complex gas dynamics and preheating
history. Indeed, it has been suggested from simulations that
preheating decreases the power law index in the region of our data
(Borgani et al.\ 2005).

\section*{ACKNOWLEDGMENT}
We thank Shai Kaspi and Sharon Sadeh for many contributing
discussions, Karl Andersson for the useful communication,
and the referee, Raphael Gavazzi, for useful comments. 
We also thank the Chandra helpdesk team Samantha Stevenson, 
Elizabeth Galle, Tara Gokas, Priya Desai, and Joan Hagler. 
We also thank Craig Gordon, Keith Arnaud, and Matthias Ehle 
for useful XSPEC tips. We acknowledge support by Israel 
Science Foundation grants 629/05 and 1218/06.

\appendix

\section{Corrleations between parameter values}

In our model-independent approach we have determined parameter values
by fitting the projected mass, gas density, and temperature. Projection
of the 3D quantities builds up correlations between their best-fit 
values in different radial bins. An illustration of the degree of such 
correlations is shown in table A1, which is the correlation matrix of 
the free parameters, taken from running four and nine free parameters 
for the total and gas mass density respectively. The correlation in the 
deduced 2D temperature is shown in table A2. The elements of the 2D 
temperature correlation matrix correspond to the 12 values of the 2D 
radii (fig.~13). In both tables the correlations are quite strong 
between adjacent elements. In the 2D temperature correlation 
matrix there are also strong correlations between elements which are not 
adjacent but are at small radii, i.e. the upper left part of the table.

\begin{table*}
\caption{The $13\times 13$ correlation matrix for $i=4$, $j= 9$ free parameters 
for the total mass and gas mass density, respectively. 
\label{correlation matrix}}
\begin{center}
\begin{tabular}{|l|c|c|c|c|c|c|c|c|c|c|c|c|c|}
\hline  
    & i=1 & i=2 & i=3 & i=4 & j=1 & j=2 & j=3 & j=4 & j=5 & j=6 & j=7 & j=8 & j=9 \\                  
\hline
i=1 & 1.000 & -0.342 & 0.197 & -0.072 & -0.026 & 0.036 & 0.034 & 0.028 & 0.040 & 0.026 & 0.026 & 0.001 & -0.006  \\  
i=2 & -0.342 & 1.000 & -0.825 & 0.273 & 0.004 & -0.005 & -0.000 & 0.010 & -0.008 & -0.051 & -0.074 & -0.005 & 0.019  \\  
i=3 & 0.197 & -0.825 & 1.000 & -0.599 & -0.002 & 0.002 & -0.000 & -0.005 & 0.030 & 0.075 & 0.091 & -0.001 & 0.004  \\  
i=4 & -0.072 & 0.273 & -0.599 & 1.000 & 0.001 & -0.002 & -0.002 & -0.003 & -0.028 & -0.045 & -0.038 & 0.018 & -0.074  \\  
j=1 & -0.026 & 0.004 & -0.002 & 0.001 & 1.000 & -0.749 & 0.323 & -0.152 & 0.057 & -0.026 & 0.010 & -0.003 & 0.002  \\  
j=2 & 0.036 & -0.005 & 0.002 & -0.002 & -0.749 & 1.000 & -0.613 & 0.266 & -0.106 & 0.047 & -0.018 & 0.006 & -0.003  \\  
j=3 & 0.034 & -0.000 & -0.000 & -0.002 & 0.323 & -0.613 & 1.000 & -0.643 & 0.238 & -0.107 & 0.042 & -0.014 & 0.007  \\  
j=4 & 0.028 & 0.010 & -0.005 & -0.003 & -0.152 & 0.266 & -0.643 & 1.000 & -0.560 & 0.234 & -0.095 & 0.031 & -0.014  \\  
j=5 & 0.040 & -0.008 & 0.030 & -0.028 & 0.057 & -0.106 & 0.238 & -0.560 & 1.000 & -0.590 & 0.223 & -0.075 & 0.038  \\  
j=6 & 0.026 & -0.051 & 0.075 & -0.045 & -0.026 & 0.047 & -0.107 & 0.234 & -0.590 & 1.000 & -0.557 & 0.174 & -0.086  \\  
j=7 & 0.026 & -0.074 & 0.091 & -0.038 & 0.010 & -0.018 & 0.042 & -0.095 & 0.223 & -0.557 & 1.000 & -0.492 & 0.226  \\  
j=8 & 0.001 & -0.005 & -0.001 & 0.018 & -0.003 & 0.006 & -0.014 & 0.031 & -0.075 & 0.174 & -0.492 & 1.000 & -0.689  \\  
j=9 & -0.006 & 0.019 & 0.004 & -0.074 & 0.002 & -0.003 & 0.007 & -0.014 & 0.038 & -0.086 & 0.226 & -0.689 & 1.000  \\  
\hline  
\end{tabular}
\end{center}
\end{table*}

\begin{table*}
\caption{The $12\times 12$ correlation matrix of the deduced 2D temperature 
for $i=4$, $j= 9$ free parameters 
for the total mass and gas mass density, respectively. 
The raw and column numbers specify radial points (see fig. 13).
\label{T2D correlation matrix}}
\begin{center}
\begin{tabular}{|c|c|c|c|c|c|c|c|c|c|c|c|}
\hline                    
1.0000 & 0.9085 & -0.6717 & 0.2874 & -0.1372 & -0.0423 & 0.0604 & -0.0008 &-0.0233 & -0.0026 & 0.0050 & -0.0005 \\
0.9085 & 1.0000 & -0.3046 & 0.1364 & -0.0854 & -0.0382 & 0.0264 & 0.0081 &-0.0008 & 0.0044 & 0.0100 & 0.0093 \\
-0.6717 & -0.3046 & 1.0000 & -0.3047 & 0.1225 & 0.0123 & -0.0773 & 0.0206 &0.0515 & 0.0162 & 0.0111 & 0.0219 \\
0.2874 & 0.1364 & -0.3047 & 1.0000 & -0.3307 & -0.1412 & 0.1199 & 0.0292 &-0.0119 & 0.0156 & 0.0363 & 0.0304 \\
-0.1372 & -0.0854 & 0.1225 & -0.3307 & 1.0000 & 0.2272 & -0.3406 & 0.0566 &0.1892 & 0.0501 & 0.0213 & 0.0626 \\
-0.0423 & -0.0382 & 0.0123 & -0.1412 & 0.2272 & 1.0000 & 0.8048 & 0.0974 &-0.1890 & 0.0403 & 0.1515 & 0.0908 \\
0.0604 & 0.0264 & -0.0773 & 0.1199 & -0.3406 & 0.8048 & 1.0000 & 0.3025 &-0.0556 & 0.0577 & 0.1344 & 0.1047 \\
-0.0008 & 0.0081 & 0.0206 & 0.0292 & 0.0566 & 0.0974 & 0.3025 & 1.0000 &0.9125 & 0.2147 & 0.0351 & 0.2234 \\
-0.0233 & -0.0008 & 0.0515 & -0.0119 & 0.1892 & -0.1890 & -0.0556 & 0.9125 &1.0000 & 0.4082 & 0.1285 & 0.2108 \\
-0.0026 & 0.0044 & 0.0162 & 0.0156 & 0.0501 & 0.0403 & 0.0577 & 0.2147 &0.4082 & 1.0000 & 0.6956 & 0.2024 \\
0.0050 & 0.0100 & 0.0111 & 0.0363 & 0.0213 & 0.1515 & 0.1344 & 0.0351 &0.1285 & 0.6956 & 1.0000 & 0.7934 \\
-0.0005 & 0.0093 & 0.0219 & 0.0304 & 0.0626 & 0.0908 & 0.1047 & 0.2234 &0.2108 & 0.2024 & 0.7934 & 1.0000 \\ 
\hline  
\end{tabular}
\end{center}
\end{table*}

\end{document}